\documentstyle[aps,prd,epsfig]{revtex}

\tightenlines
\def\ov{\overline}
\def\ra{\rightarrow}

\def\bea{\begin{eqnarray}}
\def\ena{\end{eqnarray}}
\def\bei{\begin{itemize}}
\def\eni{\end{itemize}}
\def\ben{\begin{enumerate}}
\def\enn{\end{enumerate}}

\def\mw2{M^2_W}
\def\mz2{M^2_Z}
\def\mH2{M^2_H}
\def\mh2{M^2_h}
\def\mhc2{M^2_{H^\pm}}
\def\ma2{M^2_A}
\def\mt2{m^2_t}
\def\mb2{m^2_b}

\def\lsim{\mathrel{\raise.3ex\hbox{$<$\kern-.75em\lower1ex\hbox{$\sim$}}}}
\def\gsim{\mathrel{\raise.3ex\hbox{$>$\kern-.75em\lower1ex\hbox{$\sim$}}}}

\begin{document}
\preprint{hep-ph/0103292}

\title{Very light CP-odd scalar in the Two-Higgs-Doublet Model}

\author{F. Larios \footnote{flarios@belinda.mda.cinvestav.mx}}
\address{Departamento de F\'{\i}sica Aplicada,
CINVESTAV-M\'erida, A.P. 73,  97310 M\'erida, Yucat\'an, M\'exico}
\author{G. Tavares-Velasco}
\address{Departamento de F\'{\i}sica, CINVESTAV, Apdo. Postal 14-740,
07000 M\'exico}
\author{C.-P. Yuan \footnote{yuan@pa.msu.edu}}
\address{Department of Physics and Astronomy, Michigan State
University, E. Lansing, MI 48824, USA}
\date{\today}
\maketitle

\begin{abstract}
We show that a general two-Higgs-doublet model (THDM) with a very
light CP-odd scalar ($A$) can be compatible with $\rho$ parameter,
Br($b\to s\gamma$), $R_b$, $A_b$, $(g-2)_\mu$ of muon,
Br($\Upsilon \to A \gamma$), and the direct search via the Yukawa
process at LEP. For its mass around 0.2 GeV, the muon $(g-2)_\mu$
and Br($\Upsilon \to A \gamma$) data require $\tan \beta$ to be
about 1. Consequently,  $A$ can behave like a fermiophobic CP-odd
scalar and predominantly decay into a $\gamma \gamma$ pair, which
registers in detectors of high energy collider experiments as a
single photon signature when the momentum of $A$ is large. We
compute the partial decay width of $Z \to A A A$ and the
production rate of $f \bar f \to Z A A \to Z + \/ "\gamma
\gamma"$, $f^{\prime} {\bar f} \to W^\pm A A \to W^\pm + \/
"\gamma \gamma"$ and $f \bar f \to H^+ H^- \to W^+ W^- A A \to W^+
W^- + \/ "\gamma \gamma"$ at high energy colliders such as LEP,
Tevatron, LHC, and future Linear Colliders. Other production
mechanisms of a light $A$, such as $gg \to h \to AA \to "\gamma
\gamma"$, are also discussed.

\end{abstract}

\draft

%%%%%%%%%%%%%%%%%%%%%%%%%%%%%%%%%%%%%%%%%%%%%%%%%%%%%%%%%%%%%%%%%%%%%
%12.60.Fr   Extensions of electroweak Higgs sector
%14.80.Cp   Non-standard-model Higgs bosons
%12.15.Ji   Applications of electroweak models to specific processes
%%%%%%%%%%%%%%%%%%%%%%%%%%%%%%%%%%%%%%%%%%%%%%%%%%%%%%%%%%%%%%%%%%%%%

\section{Introduction}
\label{intro}
\indent \indent

One of the main tasks of the current and future high energy
colliders is to find the Higgs boson of the Standard Model (SM), or
some other scalar particle(s), if there is any,
predicted by the extensions of the SM.
The mass spectrum of these scalars as well as their decay channels
depend on the assumed model.
Recently, the possibility of a Higgs boson decaying into a pair
of light CP-odd scalars was considered in Ref.~\cite{dobrescu}.
As pointed out in that paper, a very light CP-odd
scalar ($A$) can arise in some extensions of the SM, such as
the minimal composite Higgs model \cite{composite}, or
the next-to-minimal supersymmetric model \cite{hunter}.
An interesting aspect of a light $A$ particle is that if its mass
($M_A$) is less than twice
of the muon mass ($m_\mu$), i.e. less than about 0.2 GeV,
it can only decay into a pair of electrons ($A\to e^+e^-$)
 or photons ($A\to \gamma \gamma$).
Hence, the decay branching ratio Br($A\to \gamma \gamma$) can be
sizable. Furthermore, at high energy colliders, the light CP-odd
scalar $A$ can be produced with such a large velocity that the two
photons from its decay are highly boosted and seen by the detector
as one single-photon signature. Therefore, the production of a
pair of $AA$ is identified by the detector as a pair of photons
when each $A$ decays into its diphoton mode. The subsequent
signature as a diphoton resonance (e.g., in the case that a Higgs
boson decays into a pair of $A$) or photon cascades (e.g., in the
case that a $A$ particle is radiated from a fermion line) provides
an interesting window for the experimental search of the scalar
particles that may be responsible for the breaking of the
electroweak symmetry . In general, we expect the decay branching
ratio of $A \to \gamma \gamma$ to decrease rapidly when the
di-muon channel ($A \to \mu^+ \mu^-$) becomes available as $M_A$
increases. This is because the Yukawa coupling of
$A$-$\mu^+$-$\mu^-$ is larger than that of $A$-$e^+$-$e^-$ by the
mass ratio $(m_\mu/m_e) \sim 200$. Since we are interested in the
phenomenology of having a light $A$ decaying into a pair of
photons, we will restrict our discussion for $M_A$ less than about
$0.2$ GeV, though in principle, that is not necessary as long as
the decay branching ratio Br($A \to \gamma \gamma$) is not too
small to be observed experimentally.

A simple extension of the SM is the two-Higgs-doublet model (THDM)
\cite{hunter}, which has been extensively studied theoretically
and experimentally. For example, a constraint on the mass of the
charged Higgs boson $H^+$ in the THDM was carefully examined at
the CERN LEP as a function of its decay branching ratios into the
$\tau^+ \nu_{\tau}$ and $c {\bar s}$ modes \cite{alephc}. Studies
on searching for a light $A$ in its associated production with a
bottom quark pair were also done \cite{yukawa},
as a function of $M_A$ and $\tan
\beta$ (the ratio of the two vacuum expectation values of the
Higgs doublets in the THDM). Though some useful constraints have
been obtained by the LEP experiments,
 we show in this work that a very light $A$
(with $M_A < 2m_\mu$) is still allowed in the THDM. This low $M_A$
value, in the context of a THDM, can induce large contributions to
the $\rho$-parameter unless other parameters of the model adjust
to counteract this effect. In particular, as to be discussed
later, the masses of the charged scalar ($H^+$) and the heavy
CP-even Higgs boson ($H$) have to be approximately equal. Also,
the value of the mixing angle $\alpha$ has to be such that
$\cos(\beta - \alpha)$ becomes small in order to suppress the
Higgs boson contribution to the $\rho$-parameter.

Another important experimental data to constrain a light $A$ in
the THDM is the measurement of the muon anomalous magnetic moment.
The recent measurement at BNL \cite{bnlgm2} strongly disfavors
such a model when compared with certain theory calculation
\cite{marciano,haberdedes}. Nevertheless, other theory
calculations of the SM contributions to the muon magnetic moment
 show a better agreement with the BNL data
\cite{Yndurain}. Consequently, a light $A$ in the THDM can still
be  compatible with data though the parameter space of such a THDM
is tightly constrained.

In the next section we will examine all the relevant low energy
data (including the $(g-2)_\mu$ of muon,  Br($\Upsilon \to A
\gamma$), $\rho$-parameter, $b \to s \gamma$, $R_b$ and $A_b$) to
determine the allowed parameter space of the THDM with a very
light $A$. In section III, we consider the decay widths and the
decay branching ratios of every Higgs boson predicted in this
model. In section IV, we study the potential of the light $A$
boson as a source of the distinctive photon signal at colliders.
It can happen in either the decay mode of the neutral gauge boson
$Z\to AAA \to \/ "\gamma \gamma \gamma"$, or the production
processes of Higgs bosons, such as $f \bar f \to Z A A \to Z + \/
"\gamma \gamma"$, $f^{\prime} {\bar f} \to W^\pm A A \to W^\pm +
\/ "\gamma \gamma"$ and $f \bar f \to H^+ H^- \to W^+ W^- A A \to
W^+ W^- + \/ "\gamma \gamma"$. Section V contains our conclusion.

%%%%%%%%%%%%%%%%%%%%%%%%%%%%%%%%%%%%%%%%%%%%%%%%%%%%%%%%

\section{Constraints from low energy data}

In the THDM, the couplings of fermions to Higgs bosons are
proportional to the fermion masses. In the type-I of the THDM,
only one of the two Higgs doublets couple to fermions via Yukawa
couplings, and in the type-II of the THDM, one of the Higgs
doublets couples to the up-type fermions (with weak isospin equal
to $1/2$) and another couples to the down-type fermions (with weak
isospin equal to $-1/2$). Hence, the couplings of Higgs bosons to
fermions generally depend on the value of $\tan \beta$. In case
that the coupling of the Higgs bosons to fermions is large
and the mass of the Higgs boson is
small, the radiative correction to the low energy data can be
sensitive to the Yukawa interactions. Hence, the low energy data
can be used to impose important constraints on the masses and the
couplings of the Higgs bosons. To examine the allowed range of
$M_A$ and $\tan \beta$ in the THDM, we shall consider the
precision data on the anomalous magnetic moment of the muon, the
decay branching ratio of $\Upsilon \to A \gamma$, $b \to s
\gamma$, the $\rho$-parameter, the decay branching ratio of $Z \to
b {\bar b}$, i.e. $R_b$, and the bottom quark asymmetry $A_b$
measured at the $Z$-pole.

\subsection{Constraint on $\tan \beta$ from the $(g-2)_\mu$ of muon }

The magnetic moment of muon is defined as
\bea
\mu_\mu &=& (1+a_\mu) \frac{e \hbar}{2 m_\mu},
\nonumber \\
{\rm with} \qquad \qquad a_\mu &\equiv& \frac{(g-2)_\mu}{2} \; ,
\ena where $a_\mu$ is the muon anomalous magnetic moment, which is
induced from radiative corrections. The SM prediction includes the
QED, weak and hadronic contributions. Among them, the hadronic
contribution has the largest uncertainty, and the bulk of the
theoretical error is dominated by the hadronic vacuum polarization
(h.v.p.). There are a number of evaluations of the h.v.p.
corrections; four recent results were extensively discussed in
Ref. \cite{Yndurain}. After comparing various theory model
predictions with the precise experimental data, which is
$$ 10^{11}\times
a_\mu(\hbox{experiment})=116\,592\,023\pm140\pm60 \, ,
$$
Yndur\'ain concluded that the discrepancies between the world
averaged experiment data (\hbox{exp.}) and the theory
prediction of the SM contribution (\hbox{theo.}) are
\cite{Yndurain}:
\begin{eqnarray}
10^{11}\times\Delta a_\mu(\hbox{exp. -- theo.})=&422\pm152
(\hbox{exp.})\pm77
(\hbox{theo.})\quad \hbox{(DH)}\cr
 10^{11}\times\Delta
a_\mu(\hbox{exp. -- theo.})=&358\pm152 (\hbox{exp.})\pm112
(\hbox{theo.})\quad \hbox{(J)}\cr
10^{11}\times\Delta
 a_\mu(\hbox{exp -- theo.})=&233\pm152 (\hbox{exp.})\pm104
(\hbox{theo.})\quad \hbox{(AY)}\cr
10^{11}\times\Delta
a_\mu(\hbox{exp. -- theo.})=&119\pm152 (\hbox{exp.})\pm115
(\hbox{theo.})\quad \hbox{(CLY)}
\label{comp}
\end{eqnarray}
In the above result, DH stands for the analysis of
Ref.~\cite{davier},
J for that, as yet unpublished, of F. Jegerlehner,
quoted in Ref.~\cite{marciano},
AY indicates the result of Ref.~\cite{ay}, and
CLY is the `old' result of Ref.~\cite{cly} after being
 corrected for the new favored value of higher
order hadronic corrections \cite{Yndurain}.

In the general THDM, as well as in supersymmetric models, $a_\mu$
can receive radiative corrections at the one loop level from the
couplings of $A$, $h$ and $H$ to muons in triangle diagrams (see
Appendix~\ref{magmoment}) \cite{hunter,oneloop,gm2mssm}. As
expected, the size of the radiative corrections is proportional to
the coupling of muon to Higgs bosons. Moreover, the loop integral
reaches its maximal value when the mass of the scalar boson in the
loop becomes negligible, and diminishes as the scalar mass
increases. It was concluded in Ref. \cite{haberdedes} that a light
CP-even Higgs boson ($h$) can be responsible for the apparent
deviation of the BNL measurement of $a_\mu$ from the DH prediction
of the SM contribution, at the 90\% confidence level ( CL), in the
framework of a type-II THDM, in which the other Higgs bosons are
heavy (of the order of 100 GeV). It was found that the model
parameters have to satisfy the following requirements:
%%%%%%%%%%%%%%%%%%
\begin{eqnarray}
&& m_{\Upsilon}\lsim M_{h} \lsim 2 m_{B} \;, \nonumber \\
&& \sin(\beta - \alpha) \simeq 0 \;, \nonumber \\
&& 30\lsim \tan\beta \lsim 35 \;.
\end{eqnarray}
%%%%%%%%%%%%%%%%%%

In the case that only $A$ is light and the other Higgs bosons are
heavy, the one-loop contribution to $a_\mu$ is negative.\footnote{
In contrast, the one-loop contribution of a light $h$ to $a_\mu$ is
positive.}
Therefore, this type of new physics is strongly disfavored
according to the theory prediction of the SM contribution provided
by DH; however, it can still be compatible with the
other SM theory calculations (cf. Eq. \ref{comp}).
Furthermore, it was found in
Ref.~\cite{darwin} that a two-loop contribution to $a_\mu$ can be
sizable when $A$ is light.
As compared to the one-loop graph, a two-loop graph can contain
a heavy fermion loop. The
Yukawa coupling of the heavy fermion (with mass $m_f$) in the
second loop together with the mass insertion of the heavy fermion
will give rise to $(m_f/m_\mu)^2$ enhancement which can overcome
the extra loop suppression factor of $1/16\pi^2$.
Because the two-loop contribution can be even larger than the
one-loop contribution, the contribution from a light $A$ to
$a_\mu$ can become positive when $M_A$ is not too small. Hence, in
general, a two-loop calculation of a light $A$ contribution to the
muon magnetic moment $a_\mu$ yields a better agreement with the
experimental data than a one-loop calculation. For that reason, in
the following numerical analysis, we shall apply the two-loop
calculation presented in Ref.~\cite{darwin} to test the
compatibility of a light A THDM to the experimental data (cf.  Eq.
\ref{comp}). For completeness, we summarize the relevant formula in
Appendix~\ref{magmoment} to clarify the contributions included
in our numerical analysis.

Although we are using a two-loop calculation for our numerical analysis, it is
useful to examine a few features predicted by the one-loop calculation. Since a
one-loop contribution to $a_\mu$ from a light $A$ is always negative and the
central value of $a_\mu(\hbox{exp. -- theo.})$ is positive, the potentially large
loop contribution to $a_\mu$ has to be suppressed by a small Yukawa coupling in
order for the model to be compatible with data. For a type-II THDM, this implies a
very stringent bound on $\tan \beta $, because the coupling of muon to $A$ is
directly proportional to $\tan \beta$. (The coupling strength of
$A$-$\mu^+$-$\mu^-$ in the type-II model is $(m_{\mu}/v) \tan \beta$, where $v$ is
the weak scale, $\sim 246$ GeV.) In particular, assuming the CLY
prediction for
the SM contribution and applying the two-loop calculation to include the light $A$
contribution, we find that there exists an upper bound on $\tan \beta$
at the 95\% CL. For a 0.2 GeV pseudo-scalar,
\bea
\tan\beta < 2.6 \; .
\ena

\begin{figure}
\centerline{\hbox{ \epsfig{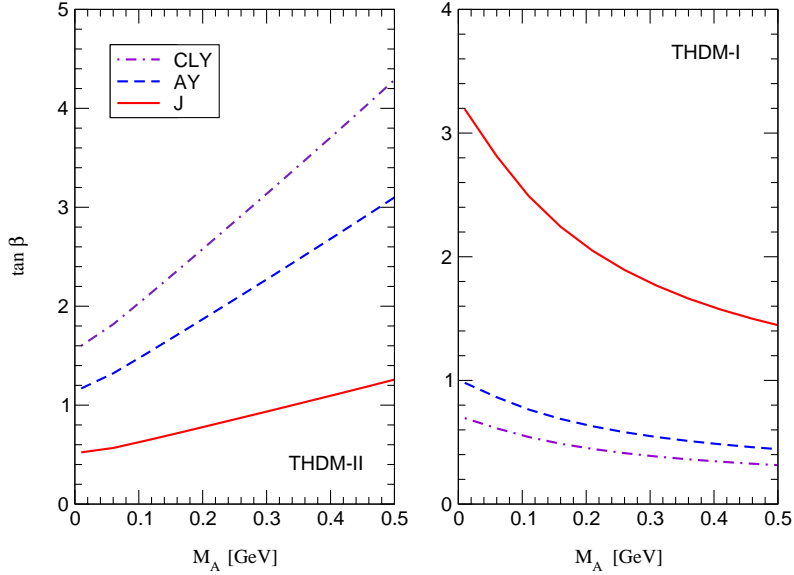}}}
\caption{The regions (below the curves for type-II and above the
curves for type-I THDM) in the $\tan\beta$ versus $M_A$ plane
allowed by the $a_\mu$ data at the 95\% CL. Three different curves
are displayed depending whether the SM prediction is given by the
CLY, AY or J calculation. There is no allowed region in this range
of parameters according to the DH calculation. (Here, a two-loop
calculation for the THDM contribution, cf.
Appendix~\ref{magmoment}, is used.) } \label{gm2tbma}
\end{figure}

\begin{figure}
\centerline{\hbox{ \epsfig{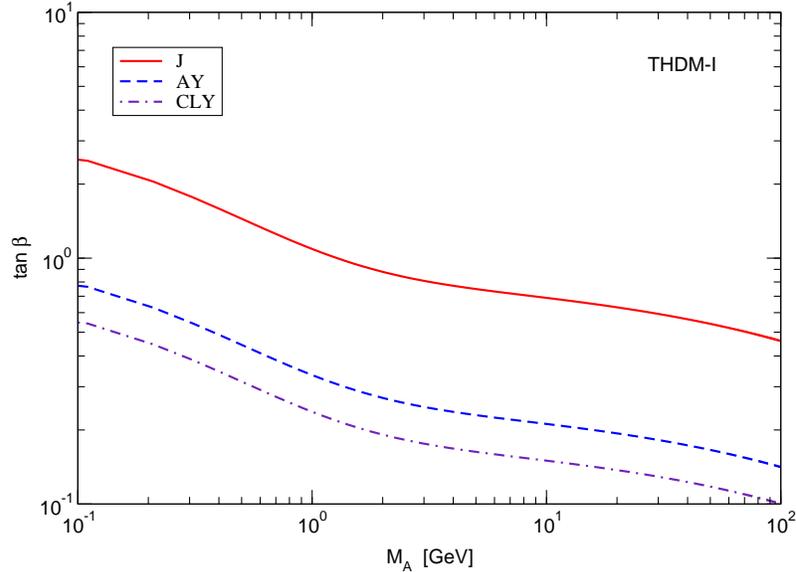}}}
\caption{The regions (above the curves) in the $\tan\beta$ versus
$M_A$ plane of a type-I THDM allowed by the $a_\mu$ data at the
95\% CL. There is no allowed region in this range of parameters
according to the DH calculation. (Here, a two-loop calculation for
the THDM contribution, cf. Appendix~\ref{magmoment}, is used.) }
\label{gm2tbma-1}
\end{figure}

\begin{figure}
\centerline{\hbox{ \epsfig{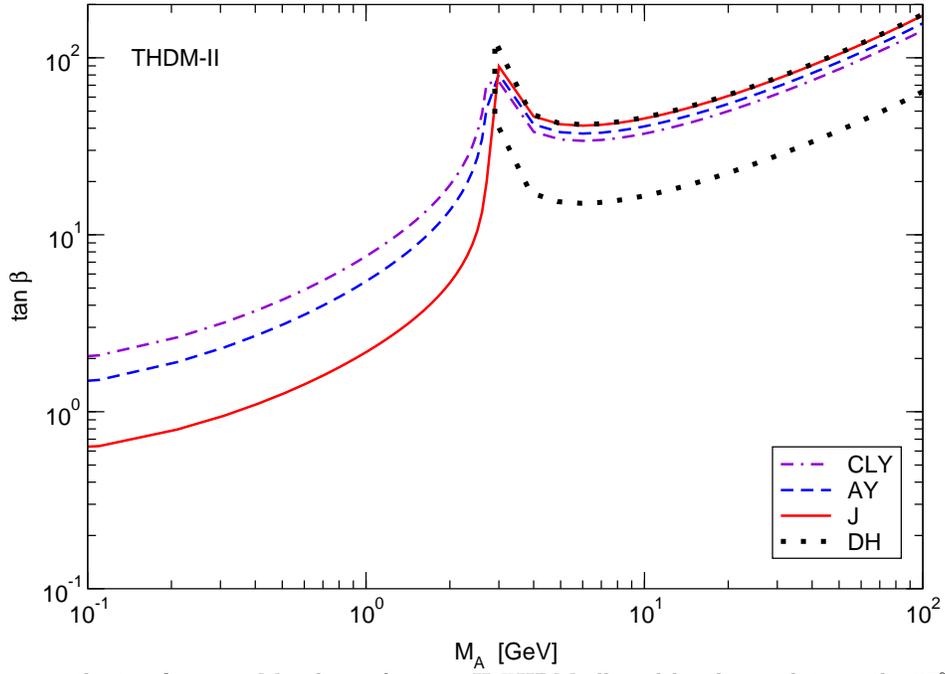}}}
\caption{The regions in the $\tan\beta$ versus $M_A$ plane of a
type-II THDM allowed by the $a_\mu$ data at the 95\% CL. The
allowed regions based on the calculations of CLY, AY and J are
below the curves. The region allowed by the DH calculation is
bounded by the dotted line. (Here, a two-loop calculation for the
THDM contribution, cf. Appendix~\ref{magmoment}, is used.) }
\label{gm2tbma-2}
\end{figure}

This new bound is stronger by a factor of two than a previous one
\cite{maria}, which was obtained from the one loop contribution,
together with the old experimental data with an error of $84\times
10^{-10}$ in the measurement of $(g-2)_\mu$. In Fig.
\ref{gm2tbma}, we show the regions in the $\tan\beta$ versus $M_A$
plane allowed by the $a_\mu$ data at the 95\% CL. Three different
curves are displayed depending whether the SM prediction is given
by the CLY, AY or J calculation. There is no allowed region
according to the DH calculation. For the type-II THDM, the allowed
regions are below the curves, for the type-I THDM, they are above
the curves.\footnote{ At the one-loop level, the pseudoscalar
contribution to $(g-2)_\mu$ in a type-I model can be obtained from
that in a type-II model after replacing $\tan \beta$ by $\cot
\beta$.} For completeness, we also show in Figs. \ref{gm2tbma-1}
and \ref{gm2tbma-2} the allowed regions in the type-I and type-II
THDM, respectively, for a wide range of $M_A$ and $\tan\beta$. In
the above figures we did not show the constraints derived from the
CUSB Collaboration search for $\Upsilon \to A \gamma$ at CESR
\cite{CUSB}. For that, we refer the readers to Ref. \cite{hunter}
which has an extensive discussion on this data to constrain a
light CP-even or CP-odd scalar in the THDM. As noted there,
various theory analyses indicated that the high order QCD
corrections to this decay rate can be large. Because of that, we
shall consider in this paper $\tan \beta$ around 1 to be
consistent with the CUSB data for a 0.2 GeV pseudoscalar.
Specifically, we shall take $\tan\beta$ to be either 0.5 or 2 in
our following discussions.

Finally, we note that a similar constraint on $\tan \beta$ can be
obtained from examining the production
of $e^- e^+ \to b {\bar b} A$ at LEP energies \cite{yukawa}.
However, as shown in Ref. \cite{maria}, the $(g-2)_\mu$ data
gives a more stringent constraint than that
obtained from direct searches of Higgs bosons via the above
Yukawa process.

\subsection{Constraint on $M_{H^+}$ from the decay of $b\to s\gamma$}

The previous $(g-2)_\mu$ analysis is only sensitive to a
light pseudoscalar $A$ when the masses of the
CP-even scalars $h$, $H$ and the charged Higgs boson $H^\pm$
are all of the order of hundred GeV. To further
constrain the parameter space of the THDM, we now turn our
attention to a low energy observable that is sensitive to a
charged Higgs boson in case that $\tan\beta$ is not
large. That is the rare decay process $b\to s \gamma$.

For the hadronic flavor changing neutral current decay process $b
\to s \gamma$, new physics effects at the weak scale can be
parameterized by the couplings (Wilson coefficients) of an
effective Hamiltonian \cite{neubert}. The only scalar contribution
to the Wilson coefficient $C_7$ in a THDM comes from the standard
penguin diagram, where the charged Higgs scalar $H^{\pm}$ couples
to top and bottom, and then to top and strange quarks.

The $H^{\pm}tb$ coupling in the type-II THDM is given by \bea
{\cal L} = \sqrt{2} V_{tb} \left \{ \frac{m_b}{v} \tan \beta \;
{\ov t_L} b_R + \frac{m_t}{v} \cot \beta \; {\ov t_R b_L} \right
\} H^+ + \rm{h.c.} \;  \label{htb} \ena where $V_{tb}$ is the
Cabbibo-Kobayashi-Maskawa (CKM) matrix element. The $H^{\pm}ts$
coupling is defined similarly with the appropriate substitution
$b\to s$. In the type-I model, the factor $\tan \beta$ is
substituted by $-\cot \beta$. For a type-II model, a value of
$\tan\beta < 1$ will induce a large ${\ov t_R} b_L H^+$ coupling,
which will strongly modify the Wilson coefficient $C_7$ and then
increase the predicted $b\to s \gamma$ rate. This can be
alleviated only if the charged $H^+$ is massive enough.

Constraints on $\tan \beta$ versus $M_{H^+}$ have been obtained by
F. Borzumati and C. Greub in the first reference
of~\cite{borzumati} for a few possible experimental upper bounds
on Br$(\bar B\to X_s \gamma)$, ranging from $3.0 \times 10^{-4}$
to $4.5 \times 10^{-4}$. Currently, the reported experimental
measurements are

\bea
{\rm Br}(\bar B \to X_s \gamma)
\times 10^4 &=& 3.15 \pm 0.35_{stat} \pm 0.32_{sys} \pm
0.26_{model}\;\; {\rm by \, CLEO\cite{cleo},} \\
{\rm Br}(\bar B \to X_s \gamma) \times 10^4 &=& 3.11 \pm
0.80_{stat} \pm
0.72_{sys} \;\; {\rm by \, ALEPH\cite{bsgaleph}, } \\
{\rm Br}(\bar B \to X_s \gamma) \times 10^4 &=& 3.36 \pm
0.53_{stat} \pm 0.42_{sys} (^{0.50}_{0.54})_{model}
\;\; {\rm by \, BELLE\cite{belle}}.
\ena

In this study, we will quote the result given in
Ref.~\cite{borzumati} for an upper bound of 4.5$\times 10^{-4}$ on
Br$(\bar B\to X_s \gamma)$. It was found that, for example, if
$\tan \beta = 0.4$, the mass of $H^+$ must be larger than about
300 GeV. On the other hand, we will not use the detailed
information on the lower bounds of $M_{H^+}$ for $\tan \beta <1$,
because in that case we would rather use the stronger bounds
obtained from examining the $b {\bar b}$ decay rate of the
$Z$-boson, i.e. $R_b$ (see next section). When $\tan \beta$ is
much greater than 1, the small effect from the ${\ov t_R} b_L H^+$
coupling is compensated by the large effect from the ${\ov t_L}
b_R H^+$ coupling; in such a way that the Br$(b\to s \gamma)$
prediction tends to stay at a certain minimal value. This minimum
only requires $M_{H^+}>165$ GeV for $\tan\beta \gg 1$. In
summary, for $\tan \beta >1$, the Br$(b\to s\gamma)$ data requires
$M_{H^+}>165$ to 200 GeV at the 95\% CL.

The situation for type-I models is different.  In this case, both
the ${\ov t_R} b_L H^+$ and ${\ov t_L} b_R H^+$ couplings decrease
for values of $\tan \beta$ greater than 1.  This does not mean
that $M_{H^+}$ can become arbitrarily small.  There is an unstable
behavior (due to the large scale dependence) for the prediction of
Br$(b\to s \gamma)$ when $M_{H^+}$ is less than about 100 GeV and
$\tan \beta \sim 1$ \cite{borzumati}. This problem of unstability
gets worse when $\tan \beta <1$.

\subsection{Constraint on $M_{H}-M_{H^+}$ from $\Delta \rho$ }

\begin{figure}
\centerline{\hbox{ \epsfig{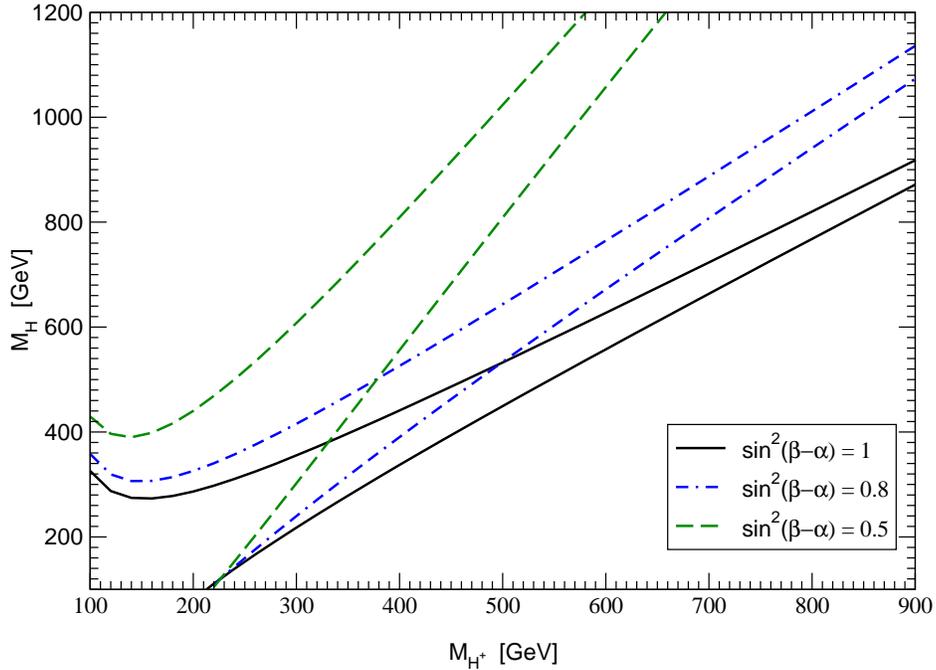}}}
\caption{Allowed regions for $M_{H^+}$ vs $M_H$ at different
values of  $\sin^2 (\beta -\alpha)$.  Regions inside the solid,
dot-dashed and dashed lines are allowed for $\sin^2 (\beta
-\alpha)=$ 1, 0.8 and 0.5, respectively.} \label{rhobounds}
\end{figure}

The effect of scalar fields ($A$, $h$, $H$ and $H^+$) on $\Delta
\rho$ have been reported for a general THDM in the literature
\cite{others,haber,logan}. When $M_A$ becomes negligible, the
contribution to $\Delta \rho$ can grow quadratically with the
masses of the other scalars (See the appendix of
Ref.~\cite{logan}). The only way to keep $\Delta \rho$ small is to
have cancellations among potentially large loop contributions from
each Higgs boson. For example, as $\sin^2 (\beta -\alpha)=1$, this
cancellation can take place between the contributions from $H^\pm$
and $H$, and requires a certain correlation between their masses.
($\alpha$ is the mixing parameter between the two CP-even Higgs
bosons $h$ and $H$.) As shown in the appendix of
Ref.~\cite{logan}, this correlation depends on the value of the
coefficient $\sin^2 (\beta -\alpha)$. In Fig.~\ref{rhobounds} we
show the different allowed regions in the $M_{H^+}$ versus $M_H$
plane for three different values of $\sin^2 (\beta -\alpha)$.
These regions or {\it bands} correspond to the 95 \% CL
limits for $\Delta \rho$ \cite{logan} with \bea -1.7 \; < \;
\Delta \rho \times 10^3 \; < \; 2.7. \ena

For simplicity, we shall assume $M_H \sim M_{H^+}$ and $\sin^2
(\beta -\alpha)=1$ for our detailed numerical analysis. It should
be noted that the allowed band for $\sin^2 (\beta -\alpha)=1$ only
depends on $M_H$ and $M_{H^+}$, whereas the allowed bands for
$\sin^2 (\beta -\alpha)<1$ do depend on $M_h$. In
Fig.~\ref{rhobounds} we have used a low value of
$M_h=110$ GeV.  If we use a higher value, for instance $M_h=130$
GeV, the bands keep the same width but slightly shift downwards
(to a somewhat smaller slope).

We have checked that within the allowed parameter space
constrained by the $\rho$-parameter, the new physics contribution
from the THDM with a light $A$ to the S-parameter is typically
small as compared to the current data: $\Delta S=-0.07 \pm 0.11$
\cite{s_data}. Using the analytical result in Ref. \cite{s_exp},
we find that $\Delta S$ approximately equals to $-0.02$ when $M_H$
and $M_{H^+}$ are about the same, and reach $-0.05$ when $M_H$ and
$M_{H^+}$ differ by a few hundred GeV. Although $\Delta S$ can
reach $-0.1$ when $M_H$ and $M_{H^+}$ differ by about a TeV, this
choice of parameters is already excluded by the $\rho$-parameter
measurement. Hence, we conclude that the current $S$-parameter
measurement does not further constrain this model.
Needless to say that the above conclusion holds for both the type-I
and type-II THDM at the one-loop order.

\subsection{Constraint on $M_{H^+}$ versus $\tan \beta$ from
$R_b$ and $A_b$}

The contributions to the $Z\to b \ov b$ hadronic decay branching
ratio ($R_b$) and the forward-backward asymmetry ($A_b$) of the
bottom quark in $Z$ decays are
given in terms of the effective $Zb\ov b$ couplings~\cite{logan}:
\bea
{\cal L} &=& \frac{e}{s_w c_w} \Big( g_L
 \; {\ov b_L} \gamma^\mu b_L \; + \; g_R
\; {\ov b_R \gamma^\mu b_R} \Big)\; Z_\mu \; ,
\\
g_L &=&   -{\frac{1}{2}}
+\frac{1}{3}s^2_w  \, +\, \delta g_L \; ,
\nonumber \\
g_R &=&
\frac{1}{3}s^2_w   \, +\, \delta g_R \; .
\nonumber
\ena
Here, $\delta g$'s contain the SM as well as the
{\it new} physics (THDM) contributions at one loop order:
\bea
\delta g_L &=& \delta g_L^{SM} + \delta g_L^{new}
\; = \; -0.4208 + \delta g_L^{new}, \\
\delta g_R &=& \delta g_R^{SM} + \delta g_R^{new} \; = \; 0.0774 +
\delta g_R^{new} \; ,
\ena
where the SM values are for $m_t=174$
GeV and $M_h=100$ GeV \footnote{In the heavy top quark mass
expansion, $R_b$ depends on $M_h$ through $\log (M_h/m_Z)$;
therefore, the dependence of $R_b$ on $M_h$ is small.}
\cite{field}. Because of the left-handed nature of the weak
interaction, the value of $g_L$ is higher (by about 5 times) than
$g_R$.

\begin{figure}
\centerline{\hbox{ \epsfig{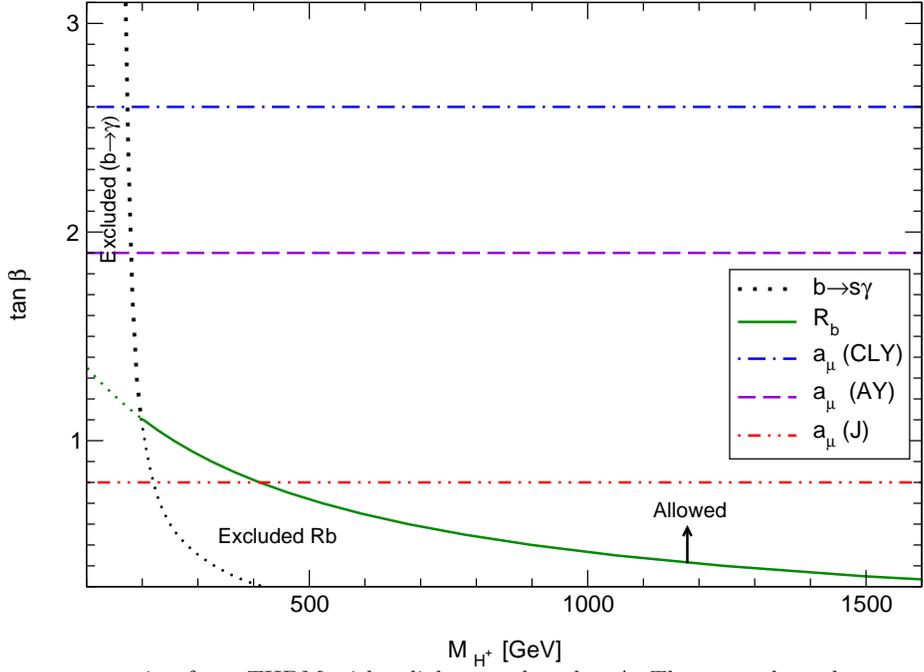}}}
\caption{Allowed parameter region for a THDM with a light
pseudoscalar $A$. The upper bounds on $\tan\beta$ derived from the
$a_\mu$ data is for a type-II THDM with $M_A=0.2$  GeV. The lower
bound from $R_b$ holds for either type-I or type-II THDM (when
$\tan\beta \sim 1$). The $b\to s\gamma$ data does not provide any
useful constraint for a type-I model when $\tan\beta > 1$.}
\label{rbounds}
\end{figure}

As shown previously, the $(g-2)_\mu$ data requires $\tan \beta$ to
be less than about 2.6 for a 0.2 GeV $M_A$.\footnote{ Unless
specified otherwise, we shall assume the theory prediction by CLY
in the following discussion. } In that case, the
contribution from the neutral Higgs boson loops to $R_b$ is
negligible and the dominant contribution comes from the charged
Higgs boson loop \cite{logan}. For a small $\tan \beta$, the $H^+$
loop contribution comes mostly from the ${\ov t_R b_L} H^+$
coupling, which is proportional to $(m_t/v)\, \cot \beta$, cf.
Eq.~(\ref{glhc}). Since the type-I and type-II models coincide on
this coupling, the bounds from the $R_b$ measurement applies to
both of the models. The $H^+$ loop contributions to $\delta g_L$
and $\delta g_R$ are given by \bea \delta g_L^{new} &=&
\frac{1}{16\pi^2} \left( \frac{m_t}{v} \cot \beta \right)^2 \left(
\frac{R}{R-1}-\frac{R\log R}{(R-1)^2} \right)
\; , \label{glhc} \\
\delta g_R^{new} &=& -\frac{1}{16\pi^2}
\left(  \frac{m_b}{v} \tan \beta \right)^2 \left(
\frac{R}{R-1}-\frac{R\log R}{(R-1)^2}\right) \; ,
\ena
with $R\equiv m_t^2/M^2_{H^+}$.
Expanding $R_b$ and $A_b$ to first order in
$\delta g_R^{new}$ and $\delta g_L^{new}$, we obtain
\bea
R_b &=& R_b^{SM} -  0.7785 \delta g_L^{new} + 0.1409
\delta g_R^{new}, \\
A_b &=& A_b^{SM} -  0.2984 \delta g_L^{new} - 1.6234 \delta
g_R^{new} \; , \ena with $R_b^{SM}=0.2158$ and $A_b^{SM}=0.935$.
As indicated in the above equation, $A_b$ depends more on $g_R$
than on $g_L$; the opposite is true for $R_b$.  For this reason,
and because the experimental uncertainty of $R_b$ is significantly
smaller than that of $A_b$, the asymmetry $A_b$ is not nearly as
effective to constrain the parameter region of the THDM than $R_b$
when $\tan\beta$ is of the order of 1. Hence, in this work, we
will only consider the constraints imposed by the $R_b$
measurement, for which the experimental limits are \cite{limits}:
\bea R^{Exp}_b &=& 0.21648 \; \pm \; 0.00075 \; . \label{rbexp}
\ena The allowed region, at the 95\% CL, in the $\tan \beta$
 versus $M_{H^+}$ plane is shown in Fig.~\ref{rbounds}, in which
the constraints imposed by the $(g-2)_\mu$ and the $b\to s\gamma$
data are also included. It is
interesting to compare our $R_b$ bound (the lower solid curve)
with the Fig. 6.1 of Ref.~\cite{logan} (same as the
Fig. 5 of Ref.~\cite{haber}). There, the experimental central
value used for $R_b$ was 0.21680, with a smaller ($1\sigma$) error
of $\pm 0.00073$.
Since the new $R^{Exp}_b$ is only $0.81\sigma$
above the SM value that we use here, the bounds on $M_{H^+}$
become less stringent\footnote{For our $R_b$ analysis, we used the
same input parameters as those given in Appendix H of
Ref.~\cite{logan}, but with an updated value for $R^{Exp}_b$.}.

In summary, Fig.~\ref{rbounds} shows the allowed region in the
$\tan \beta$ versus $M_{H^+}$ plane for the type-I and type-II
models with a light (0.2 GeV) CP-odd scalar $A$.
The $(g-2)_\mu$ data imposes an upper bound on $\tan\beta$ for
a type-II THDM, and the value of this upper bound depends on the
model of theory calculations.
Note that there is no DH curve in the figure. This is because
a THDM with a very light $A$ cannot be compatible with data
according to the DH calculation of the SM contribution to $a_\mu$.
Similarly, the lower bound on $\tan\beta$ imposed by the
$(g-2)_\mu$ data for a type-I THDM can be
easily obtained from Fig.~\ref{gm2tbma}.
The lower bound from $R_b$ holds for either type-I or type-II THDM,
and is not sensitive to the actual value of the mass of the light
$A$, because for $\tan\beta \sim 1$
the charged Higgs boson loop dominates. A similar
conclusion also holds for the constraint imposed by the
$b\to s\gamma$ data.
Furthermore, the $b\to s\gamma$ data does not provide a useful
constraint for a type-I model when $\tan\beta > 1$. For clarity,
we summarize the constraints from the relevant low energy data in
Table~\ref{allconstraints} for each type of THDM. Although in the
type-II THDM, the value of $\tan \beta$ is bounded from above to
be less than about 2.6, it can take on any value above the lower
bound imposed by the $R_b$ data. On the other hand, the value of
$\tan \beta$ in the type-I model cannot be too large because for a
very large value of $\tan \beta$, the decay width of the lighter
CP-even Higgs boson ($h$) can become as large as its mass, so that
the model ceases to be a valid effective theory.

%\newpage
%\vspace{1cm}

%\vspace{2cm}

 In the next section, we shall discuss the decay and production
of the Higgs bosons in the THDM with a light $A$. For simplicity,
we shall only discuss the event rates predicted by a type-II
model, and the value of $\tan\beta$ is taken to be either 0.5 or 2
to be consistent with all the low energy data discussed in this
section. As shown in Fig. \ref{rbounds}, in a type-II THDM with a
light $A$, the masses of the Higgs bosons other than the light
CP-even Higgs boson $h$ have to be around 1 TeV when
$\tan\beta=0.5$. Hence, its phenomenology can be very different
from the model with $\tan\beta=2$, in which a few hundred GeV
heavy Higgs boson $H$ and charged Higgs boson $H^\pm$ are allowed.

\section{Decay Branching Ratios of Higgs Bosons}

In this section, we shall examine the decay branching ratios of
the Higgs bosons predicted by the type-II THDMs with a
very light CP-odd scalar $A$. To be consistent with the
$\rho$-parameter analysis, the mixing angle $\alpha$ as well as
the Higgs boson masses $M_{H^+}$ and $M_{H}$ have to be
correlatively constrained, cf. Fig. \ref{rhobounds}. Without
losing generality, in this section we shall assume $\sin (\beta
-\alpha)=1$ to simplify our discussion.

As noted in the previous sections, a CP-odd scalar $A$, with its
mass around 0.2 GeV, can only decay into a $e^-e^+$ pair
or a photon pair. Though the decay process $A\to \gamma \gamma$
can only occur at loop level, its partial decay width may compete
with the tree level process $A \to e^-e^+$. This is because the
mass of the electron is very tiny as compared to the electroweak
scale $v$, and the partial decay width of $A \to e^-e^+$ is
suppressed by ($m_e/v)^2$.

To clarify our point, we note that for $\alpha=\beta-\pi/2$, the
couplings of Higgs bosons and fermions in the THDM are given by
\bea {\cal L}_{\ov f f S} &=-& \frac{m_f}{v} {\ov f} f h \; -\; X
\frac{m_d}{v} \ov d d H \; +\; Y\frac{m_u}{v}
\ov u u H \nonumber \\
&+&  i X \frac{m_d}{v} \ov d \gamma^5 d A \; +\;
i Y \frac{m_u}{v} \ov u \gamma^5 u A ,
\label{eq:ffs_coup}
\ena
where $X=\tan \beta$, $Y=\cot \beta$ for the type-II model,
and $X=-\cot \beta$, $Y=\cot \beta$ for the type-I model.
(Here, $f$ stands for any fermion, $u$ for an up-type fermion,
and $d$ for an down-type fermion.)
Therefore,
the partial decay widths of Higgs bosons into fermion pairs are:
\bea
\Gamma_{{S_i}\to f{\bar f}} &=& \frac{N_c}{8\pi} C_{{\ov f}fS}^2
M_{S_i} \left( 1-4\frac{m^2_f}{M^2_{S_i}} \right)^{3/2},
\, {\rm with} \, {S_i}=h \,\, \/ {\rm or} \/ \,\, H ,
\\
\Gamma_{A\to f{\bar f}} &=& \frac{N_c}{8\pi} C_{{\ov f}fA}^2
M_A \left( 1-4\frac{m^2_f}{M^2_A} \right)^{1/2},
\ena
where $C_{{\ov f}f{S_i}}=\frac{m_f}{v}\times${\small
(1, $X$, or $Y$ )} is the coupling defined in ${\cal L}_{\ov f f S}$,
and $N_c$ is the color factor (which is 3 for quarks and 1 for
leptons).

The partial decay width of $A \to \gamma \gamma$ at the one-loop
order arises from fermion loop contributions, which yield \cite{hunter}:

\begin{equation}
\Gamma\left(A\to \gamma \gamma \right)=\frac{\alpha^2 g^2}{256
\pi^3}\frac{M_A^3}{m_W^2}\left|H\right|^2,
\end{equation}

\begin{equation}
H=\sum_f N^f_c\, Q_f^2\, C_f\,
\tau_f\, F\left(\tau_f\right)
\end{equation}

\noindent where $\tau_f=(2 m_f/M_A)^2$, $N^f_c=1\,(3)$ for leptons (quarks), $Q_f$
and $m_f$ are the electric charge (in units of $e$) and the mass of fermion,
respectively. Also,  $C_f= \cot \beta\,(\tan \beta)$ for up-type quarks (charged
leptons and down-type quarks) in a type-II THDM. Furthermore, $F(x)$ is given by

\begin{equation}
\label{Ffun}
F(x)=\left\{\begin{array}{ll}
\left(\arcsin \frac{1}{\sqrt x}\right)^2 & \qquad {\rm for} \;  x\geq 1 \\
-\left({\mathrm{arccosh}}\frac{1}{\sqrt x}-\,\frac{i \,\pi}{2}\right)^2
&\qquad {\rm for} \; x<1,
\end{array}
\right.
\end{equation}

\begin{figure}
\centerline{\hbox{ \epsfig{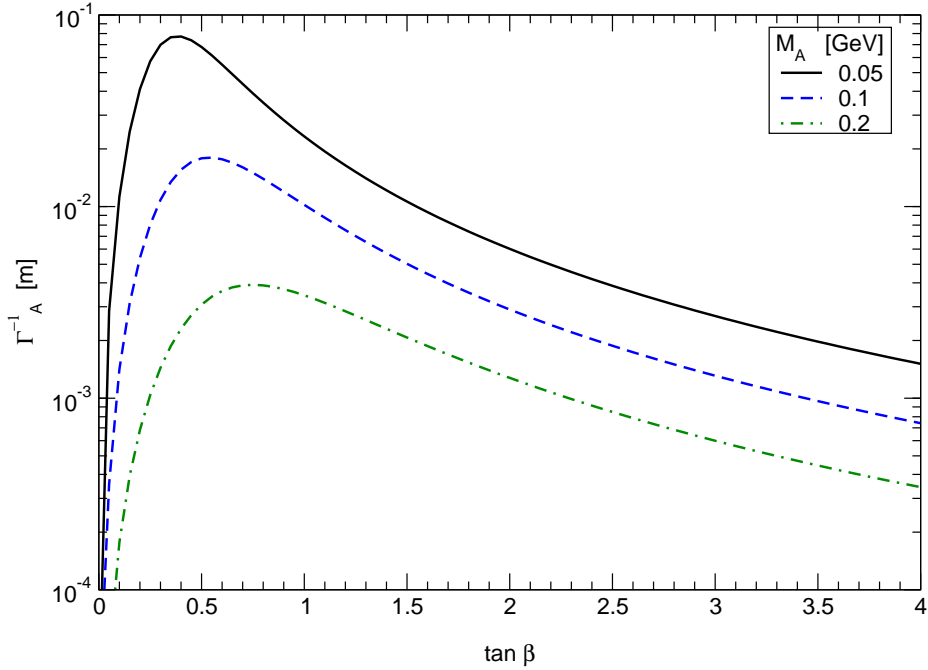}}}
\caption{Life-time of $A$ for $M_A= 0.05, 0.1$ and $0.2$ GeV. The
units have been converted from GeV$^{-1}$ to meters. The decay
length of $A$ can be obtained after multiplying $1/\Gamma_A$ by
$|{\vec p}|/M_A$, where ${\vec p}$ is the momentum of $A$. }
\label{invw}
\end{figure}

\begin{figure}
\centerline{\hbox{ \epsfig{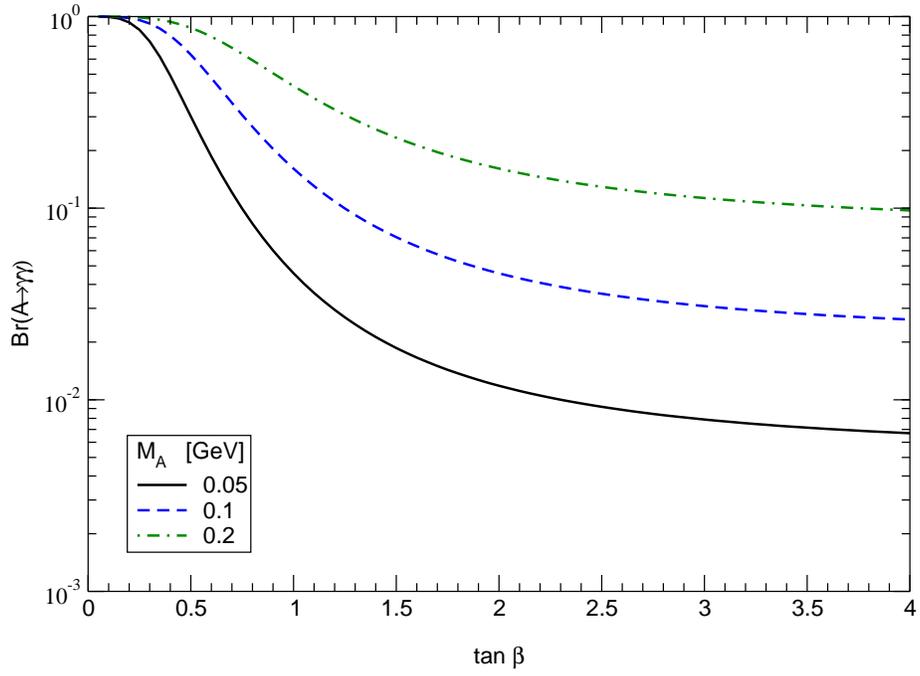}}}
\caption{Branching ratio of $A\to \gamma \gamma$ for $M_A=$ 0.05,
0.1, and 0.2 GeV.} \label{abran}
\end{figure}

In Figs.~\ref{invw} and \ref{abran}, we show the life-time
(multiplied by the speed of light) of $A$ and its partial decay
branching ratio Br($A \to \gamma \gamma$), as a function of
$\tan \beta$ for various $M_A$ values.
As indicated, the typical life-time of a light CP-odd scalar $A$
(with mass around 0.2 GeV and $\tan\beta$ around 1)
is about $10^{-3}$ meter, so the decay
length of a 50 GeV $A$ boson is about 0.25 meter. This
unique feature of a light scalar $A$ boson can be used to
improve identifying such an event experimentally.
For $\tan \beta \sim 1$, about half of the time, the light $A$ can
decay into a photon pair, other than a $e^-e^+$ pair.
Because $A$ is usually produced with a large velocity in collider
experiments, the two decay photons will be largely boosted and seen
by the detector (the electromagnetic calorimeter) as if it were a
single-photon signal.

To discuss the decay branching ratios of the other Higgs bosons,
we need to specify all the parameters in the scalar sector of the
THDM Lagrangian. In a CP-conserving THDM Lagrangian with natural
flavor conservation (ensured by the discrete symmetry of $\phi_1
\to \phi_1$ and $\phi_2 \to - \phi_2$), there are eight parameters
in its Higgs sector \cite{logan}. They are $m_1$, $m_2$,
$\lambda_{1,2,3,4,5}$, and $\mu_{12}$, or equivalently, $M_h$,
$M_H$, $M_{H^+}$, $M_A$, $\alpha$, $\tan \beta$, $v$, and
$\mu_{12}$. Of these eight free parameters, seven have been
addressed in the previous
section: four Higgs boson masses, two mixing angles ($\alpha$ and
$\beta$) and the vacuum expectation value $v$. There is yet
another free parameter: the soft breaking $\mu_{12}$ term that so
far has not been constrained.\footnote{ The parameter $\mu_{12}$
is defined through the interaction term  $-\mu_{12}^2 {\cal R} \{
\phi_1^{\dagger} \phi_2 \}$, which softly breaks the discrete
symmetry of $\phi_1$ and $\phi_2$. It is given by $2\lambda_5 v_1
v_2 \cos \xi$ in Ref.~\cite{hunter}; it is sometimes written as
$m_3$, see Ref.~\cite{shinya} for example.} This is because up to
the one loop order, $\mu_{12}$ does not contribute to the low
energy data discussed above. With the assumption $\alpha =\beta -
\pi/2$ we can write the $hAA$, $HAA$, $HAZ$ and $H^+AW$ couplings
as follows: \bea {\cal L} &=&  \frac{1}{2} \lambda_h hAA \; +\;
\frac{1}{2} \lambda_H HAA \; + \label{hcouplings} \\
&{\,}& \lambda_{HAZ} \; Z^\mu
(H\partial_\mu A - A\partial_\mu H) \; +\;
\lambda_{H^+AW} \; \left[W^-_\mu
(H^+\partial^\mu A - A\partial^\mu H^+) + \rm{h.c.}\right],
\nonumber
\ena
\noindent
where the coupling constants are given by
\bea
\lambda_h &=& \frac{(2M^2 - M^2_h)}{v}, \nonumber \\
\lambda_H &=& 2\frac{M_H^2 - M^2}{v \tan 2\beta},
\nonumber \\
\lambda_{HAZ} &=& \frac{-e}{2 s_W c_W},
\nonumber \\
\lambda_{H^+AW} &=& \frac{-e}{2 s_W},
\nonumber
%\label{hcoup}
\ena
with $M^2 = {\mu^2_{12}}/({\sin \beta \,\cos \beta})$.
For a very light $A$, both the total width and the decay branching
ratios of the other Higgs bosons in the model can be strongly
modified to differ from the usual predictions of
the THDM (with $M_A$ at the weak scale $v$).

\begin{figure}
\centerline{\hbox{ \epsfig{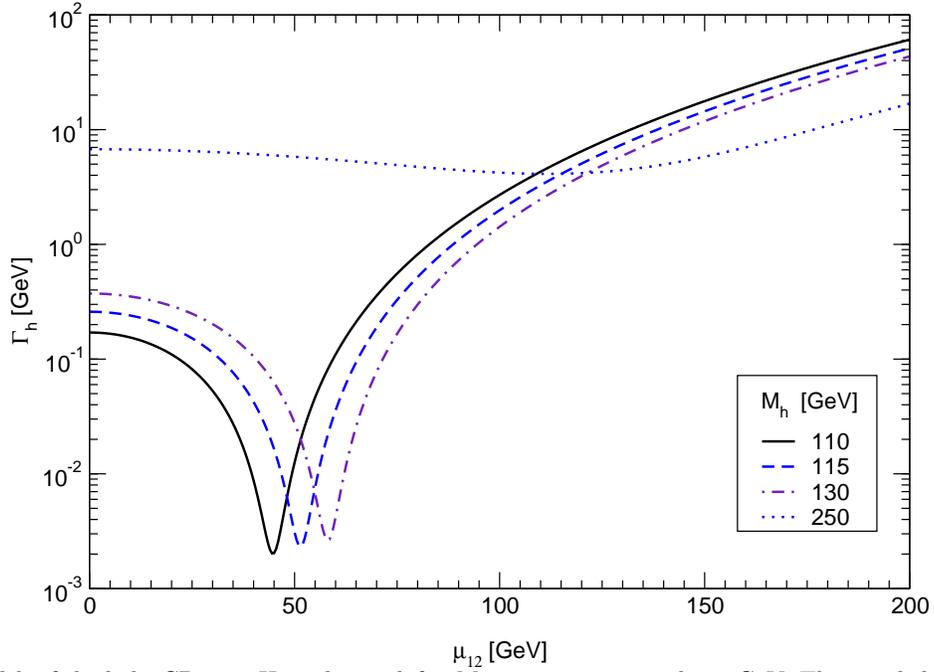}}}
\caption{Total width of the light CP-even Higgs boson $h$ for
$M_h=$100,115,130 and 250 GeV. This result holds for $\tan
\beta=0.5$ or 2. } \label{hwidth}
\end{figure}

At tree level, the partial decay width of
$h \to AA$ (or $H \to AA$) is given by
\bea
\Gamma_{(H)h\to A A} =
\frac{\lambda^2_{(H)h}}{32\pi M_{(H)h}}.
\ena
It turns out that $h\to AA$ is always the dominant (more than
90\%) decay channel of $h$ except when $2M^2 \simeq M^2_h$, in
which case the $hAA$ coupling is diminishing. When the parameter
$\mu_{12}$ increases, the decay width
of the light Higgs boson can become large.
For instance, for $M_h=110$ GeV and $\mu_{12}=200$ GeV,
$\Gamma_{h\to A A}=34$ GeV and 54 GeV for $\tan \beta =1$ and 2,
respectively. Hence, in order for the considered model to be a
valid effective theory we shall restrict the range of the
parameter $\mu_{12}$ so that the decay width of any Higgs boson
should not be as large as its mass.  For that reason, in the rest
of this study, we shall consider the range of $\mu_{12}$ to be
$0<\mu_{12}<200$ GeV.\footnote{For simplicity, we assume
$\mu_{12}$ to be a positive value, though the same conclusion also
hold for a negative value of $\mu_{12}$.} In Fig.~\ref{hwidth}, we
show the total decay width ($\Gamma_h$) of $h$ as a function of
$\mu_{12}$ for a few values of $M_h$. Here, we take the value of
$\tan \beta$ to be 0.5.
The similar figure for $\tan \beta=2$ is
identical to that for $\tan \beta=0.5$. This is because we are
considering $\alpha=\beta-\pi/2$ and
the partial decay width of $h \to AA$ is unchanged after
replacing $\tan\beta$ by $\cot\beta$.

For $M_h < 2 m_W$, where $m_W$ is the mass of the
$W^\pm$ gauge boson, $h \to b {\bar b}$ is the sub-leading decay
mode except when $M^2$ is in the vicinity of $M^2_h/2$. For $M_h >
2 m_W$, the other decay modes (e.g., $h \to W^+W^-, ZZ, t {\bar
t}$) can open, and in that case, the $AA$ mode is usually not the
dominant decay mode. Since we are interested in the $A \to \gamma
\gamma$ signal, we shall restrict our attention to values of $M_h$
for which the decay branching ratio of $h \to AA$ can be sizable.
To give a few examples,  we show in Fig.~\ref{lightbr} the
branching ratios of $h \to AA \;, b {\bar b}$ for
$M_h=100,115,130$ GeV with $\tan \beta=0.5$.
(Again, the similar figure for $\tan \beta=2$ is
identical to that for $\tan \beta=0.5$.)

In addition to the $AA$ channel, a heavy CP-even Higgs boson $H$
can also decay into the $AZ$ mode with a sizable branching ratio.
The partial decay width for $H\to AZ$ is
\bea \Gamma_{H\to AZ} &=& \frac{\lambda_{HAZ}^2 M_H}{16\pi}
\frac{(1-R_{ZH})^3}{R_{ZH}} \; , \label{HAZwid}
\ena
\noindent with $R_{ZH}=m_Z^2/M_H^2$. (We note that $\Gamma_{H\to
AZ}$ does not depend on $\mu_{12}$.) In Fig.~\ref{heavywidth}, we
show the total decay width of $H$ as a function of $\mu_{12}$ for
a few values of $M_H$. Its decay branching ratios into the $AA$
and $AZ$ modes are shown in Fig.~\ref{heavybran2} for various
$M_H$ values with $\tan \beta=2$.
To study the $\tan \beta$
dependence, we also show in Fig.~\ref{heavybratan} the branching
ratios of $H \to AA \;, ZA$ for $\mu_{12}=100$ GeV. In this case,
the sub-leading decay modes are $H \ra b {\bar b} \; ,t {\bar t}$,
etc.
For $M_H$ less than twice of the top quark mass, the
curves in Fig.~\ref{heavybratan} are almost symmetric with
respect to $\tan\beta=1$. This is again because we have set
$\alpha=\beta-\pi/2$ and $\tan\beta$ is of order 1.

The decay branching ratios of $H^+$ are also largely altered in the
THDM with a very light $A$, because the $H^+\to AW^+$ channel becomes
available. At the Born level,
the partial decay width $\Gamma_{H^+\to AW^+}$ can be
calculated using the formula given in Eq. (\ref{HAZwid}) after
substituting $H\to H^+$ and $Z\to W^+$.\footnote{
The one loop corrections to $\Gamma_{H^+\to AW^+}$ have been
calculated in Ref.~\cite{akeroyd}; they can modify the tree level
width up to a few percent for the very low values of $M_A$
considered here.
}
In Fig.~\ref{hcbraii}, we show the dominant branching ratios of $H^+$
for the type-II model, in which the $b \to s \gamma$ data requires
$M_{H^+} > 165$ GeV.

%%%%%%%%%%%%%%%%%%%%%%%%%%%%%%%%%%%%%%%%%%%%%%%%%%%%%%%
\section{Probing a Light $A$ at High Energy Colliders}

In this section, we discuss the potential of the present and future
high energy colliders for detecting a light CP-odd scalar $A$ to test
our scenario of the THDM.

The exciting feature of a light CP-odd scalar $A$ is that
the light CP-even Higgs boson $h$ in the THDM can have a large
decay branching ratio into the $AA$ mode, and each $A$ subsequently
decays into a photon pair.
For $h$ with mass around $100$ GeV, the decay particle $A$ with
mass around $0.2$ GeV will be significantly boosted, so that the
two decay photons from $A$ are produced almost collinearly in the
detector.  When these two almost collinear photons cannot be
resolved in the electromagnetic calorimeter, they will be
reconstructed as a single photon. (The angular resolution for
discriminating two photons in a typical detector will require
$M_h/M_A < 40$.)  As a result, the decay process
$h \to A A$ will appear in the detector as a diphoton signature, and
$Z \to A A A$ as a triphoton signature, etc.
Similarly, the final state of the production process
$e^- e^+ \to Z A A$ will appear as a $Z + 2 \gamma$ signature.

In the following, we shall discuss in details the prediction of our
scenario of the THDM on the decay branching ratio of
$Z \to A A A$ and the production rate of $ Z A A$ at the CERN LEP,
Fermilab Tevatron, CERN Large Hadron Collider (LHC) and the
future Linear Collider (LC). Other relevant production processes at
the Tevatron and the LHC will also be discussed.
Without losing generality, we again assume $\alpha=\beta-\pi/2$,
motivated by the $\rho$-parameter constraint.
Furthermore, we again take $M_A$ to be $0.2$ GeV such that the decay
branching ration of $A \to \gamma \gamma$ is large and the
life-time of $A$ is short enough to be detected inside
detectors of high energy collider experiments.

\subsection{The Decay Branching Ratio of $Z \to A A A$}

In the THDM, the $Z$ boson can decay into the $AAA$ mode via the
two tree-level Feynman diagrams shown in Fig.~\ref{treezaaa}.
Since in the case of $\sin(\beta-\alpha)=1$, the coupling of
$Z$-$A$-$h$ vanishes, only the diagram with the coupling of
$Z$-$A$-$H$ survives.  For $M_H > m_Z$, due to the suppression
factor from the three-body phase space, the partial decay width of
$Z \to A A A$, denoted as $\Gamma(Z \to A A A)$, at the tree
level, is small. For example, for $M_H=180$ GeV, $\mu_{12}=100$
GeV, and $\tan \beta=0.5$ or 2, $\Gamma(Z \to A A A) = 1.32 \times
10^{-8}$ GeV, which implies the decay branching ratio of the $Z$
boson into three isolated photons (as identified by detectors) is
of the order of $10^{-8}$ and $10^{-11}$ for $\tan\beta=0.5$ and
2, respectively. (When $M_A=0.2$ GeV the decay branching ratio
Br($A \to \gamma \gamma$) is 0.87 and 0.2 for $\tan\beta=0.5$ and
2, respectively.) For a much heavier $H$, this tree level decay
rate becomes negligible. In that case, a loop induced decay
process might be more important. In Appendix~\ref{zaaabox}, we
show the fermionic loop corrections to the decay width of $Z \to A
A A$, assuming that the other heavy Higgs bosons are so heavy that
they decouple from the low energy data. We find that in general,
this decay width is small unless the value of $\tan \beta$ is very
large.

\subsection{The Production Rate of $Z A A$ at LEP and LC}

A light $A$ could have been produced copiously at LEP-1 and LEP-2
experiments via the Yukawa process $e^- e^+ \to b {\bar b} A$
\cite{yukawa}. By searching for a light CP-odd Higgs boson in the
associate production of the bottom quark pair, LEP experiments
were able to exclude a range of $M_A$ as a function of $\tan
\beta$, when $A$ decays into a fermion (lepton or quark) pair. On
the contrary, the decay mode we are considering in this paper is
$A \to \gamma \gamma$ which will likely register into detectors as
a single-photon signal. Given the information on the decay
branching ratio ${\rm Br}(A \to \gamma \gamma)$, it is possible to
further constrain this model by examining the $b {\bar b} \gamma$
events. However, as discussed in the previous sections, the
$(g-2)_\mu$ data has already constrained $\tan \beta$ to be small
(less than about 2.6), so the production rate of $e^- e^+ \to b
{\bar b} A$ is not expected to be large at LEP. Here, we would
like to consider another possible signal of a light $A$ at LEP
experiments, i.e. via the production process $e^- e^+ \to Z A A$.

The Feynman diagrams that contribute to the scattering process
$e^- e^+ \to Z A A$ at the Born level are shown in
Fig.~\ref{diagffzaa}. With $\alpha=\beta - \pi/2$, the tree level
couplings $Z$-$Z$-$H$ and $Z$-$A$-$h$ vanish. Since by its
definition, the mass of $H$ is larger than that of $h$, the
production cross section is dominated by the diagram (a) with $h$
produced at resonance when $M_h < \sqrt{S}-m_Z$, where $\sqrt{S}$
is the center-of-mass energy of the $e^-e^+$ collider. Though the
above observation is generally true, it is possible to have the
value of $\mu_{12}$ such that the coupling of $h$-$A$-$A$ (i.e.
$\lambda_h$ in Eq.~(\ref{hcouplings}) ) becomes so small that the
production rate is instead dominated by the diagram with a $H$
boson resonance. In that case, the event signature is to have a
resonance structure in the invariant mass distribution of the $Z$
boson and one of the $A$ particles (i.e. one of the two photons
observed by the detector), provided $M_H < \sqrt{S}-M_A$.
Obviously, when $M_H > \sqrt{S}-M_A$, we do not expect any
enhancement from the resonance structure, and the cross section
becomes small. However, for a large value of $\mu_{12}$, the width
of $h$ can become so large (cf. Fig. \ref{hwidth}) that even for
$M_h > \sqrt{S}-m_Z$, the production rate of $ZAA$ can still be
sizable. The same effect also holds when the width of $H$ becomes
large. In Figs. \ref{sigmazaa} and \ref{sigmazaa05}, we show the
production cross section for $e^- e^+ \to Z A A$ as a function of
$\mu_{12}$ at the LEP and the LC for a few values of $M_h$ and
$M_H$ with $\tan\beta=2$ and $\tan\beta=0.5$, respectively. (For
completeness, we also give its squared amplitude in
Appendix~\ref{ampeezaa}.)

\begin{figure}
\centerline{\hbox{ \epsfig{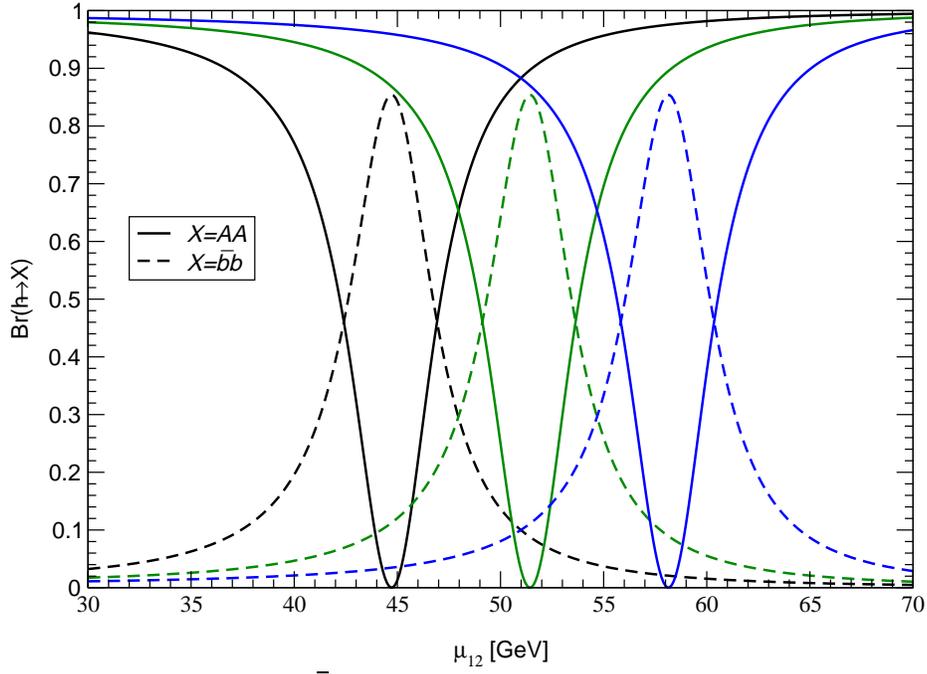}}}
\caption{Branching ratio of $h \to AA$ and $h \to b \overline b$,
as a function of $\mu_{12}$ for different $M_h$ values. From left
to right: $M_h=$100, 115 and 130 GeV. This result holds for $\tan
\beta=0.5$ or 2.} \label{lightbr}
\end{figure}

\begin{figure}
\centerline{\hbox{ \epsfig{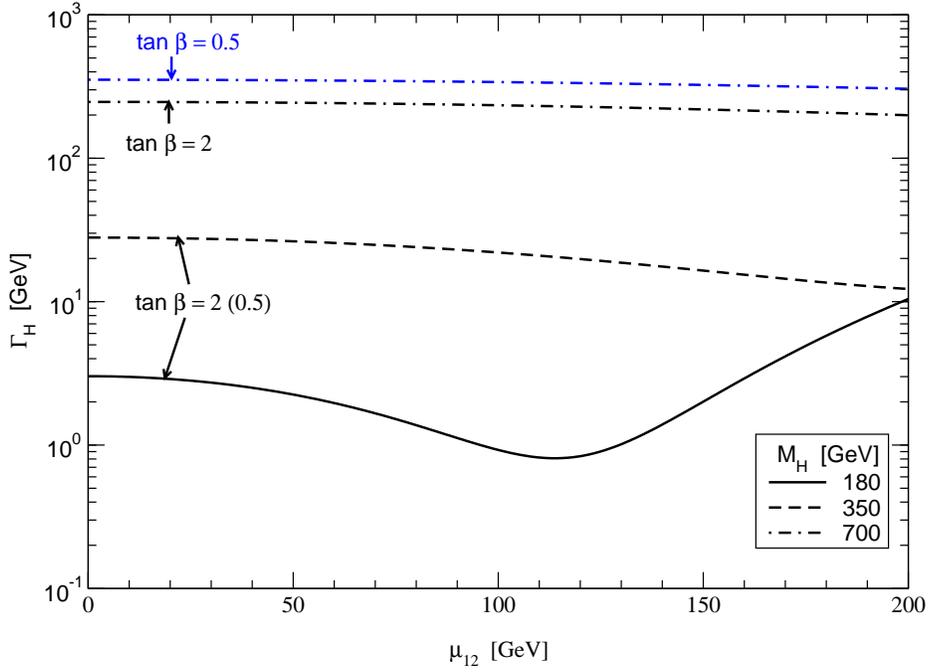}}}
\caption{Total width of $H$ as a function of $\mu_{12}$ with $\tan
\beta =0.5$ or 2, for $M_H=$ 180, 350 and 700 GeV.}
\label{heavywidth}
\end{figure}

\begin{figure}
\centerline{\hbox{ \epsfig{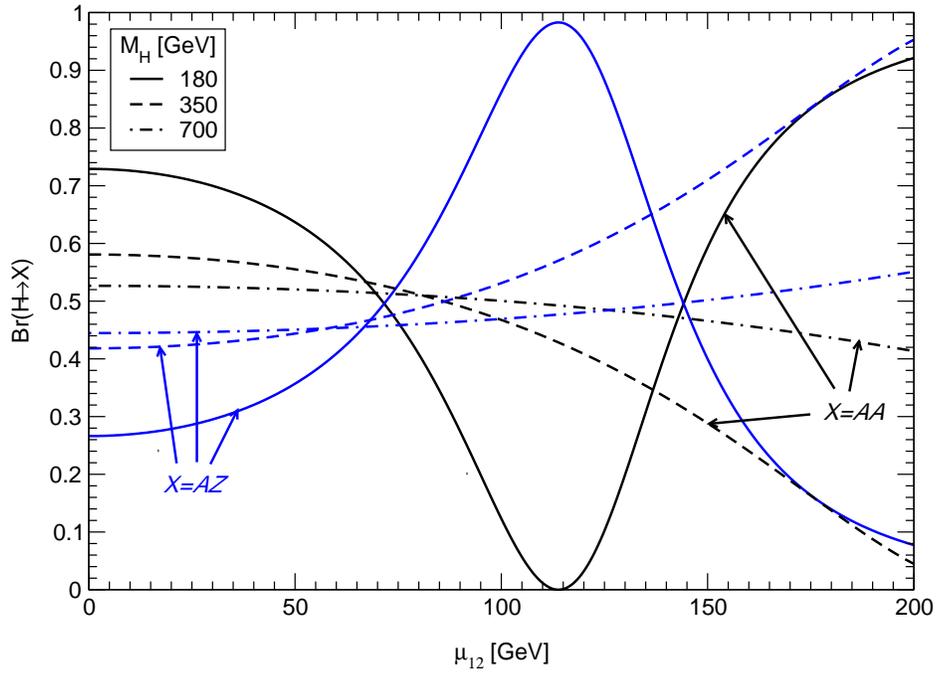}}}
\caption{Branching ratios of $H\to AA$ and  $AZ$ with $\tan
\beta=2$, for $M_H=$ 180, 350 and 700 GeV.} \label{heavybran2}
\end{figure}

\begin{figure}
\centerline{\hbox{ \epsfig{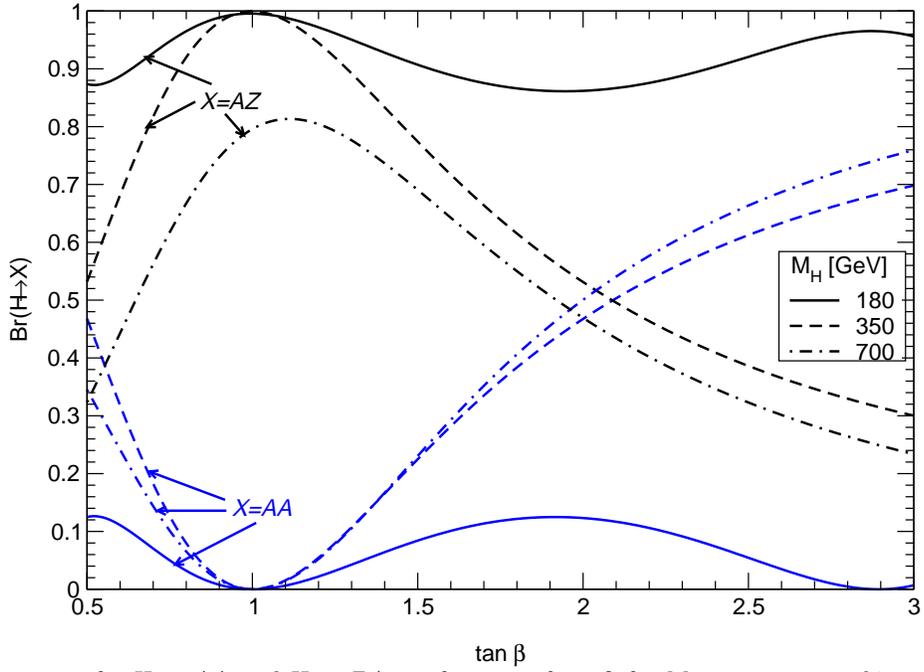}}}
\caption{Branching ratios for $H\to AA$, and $H\to ZA$ as a
function of $\tan \beta$, for $M_H=180$, 350 and 700 GeV with
$\mu_{12}=100$ GeV.} \label{heavybratan}
\end{figure}

\begin{figure}
\centerline{\hbox{ \epsfig{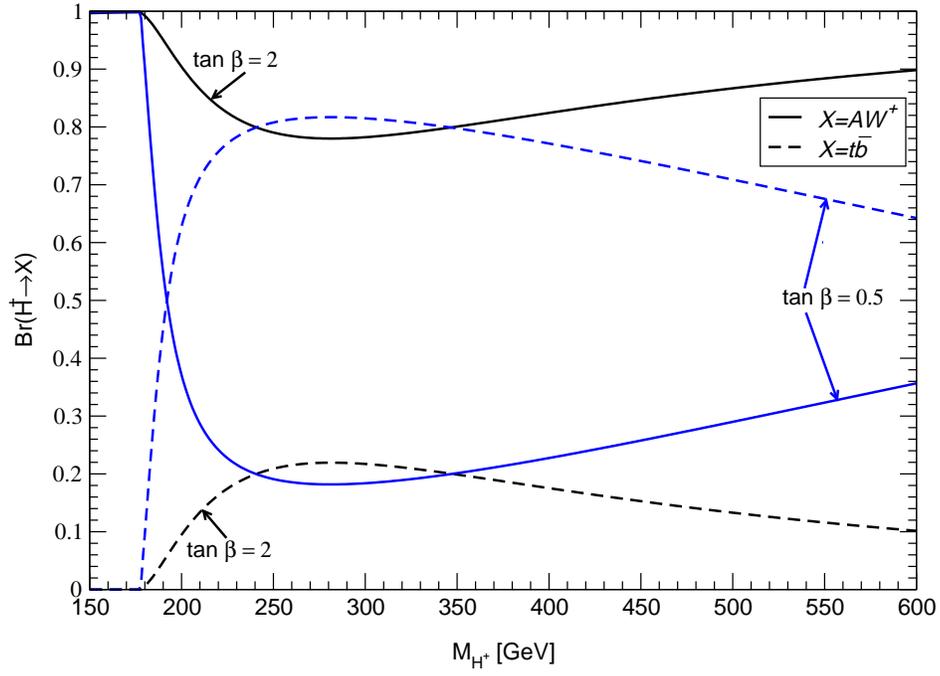}}}
\caption{The $H^+\to AW^+$ and $H^+\to t\ov b$ branching ratios
for two values of $\tan \beta$ in the type-II THDM.}
\label{hcbraii}
\end{figure}

\begin{figure}
\centerline{\hbox{ \epsfig{figure=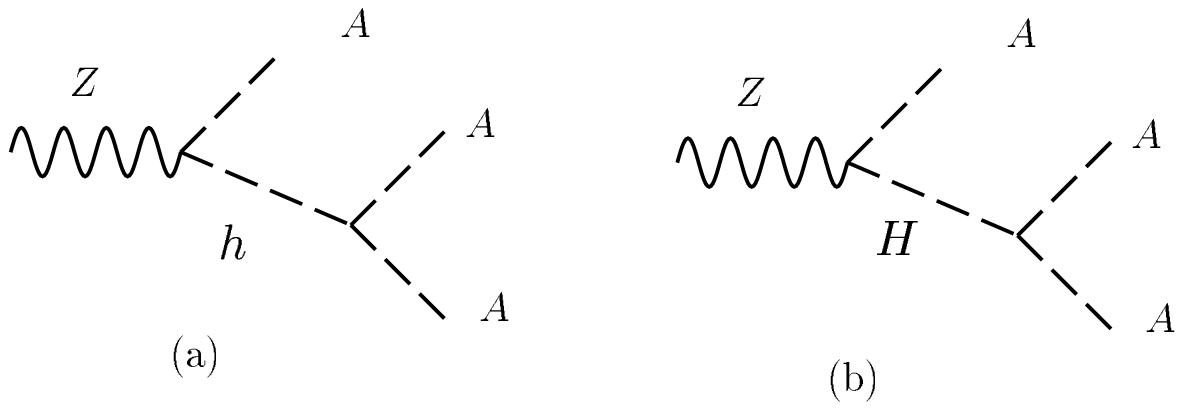,height=1.5in}}}
\caption{Tree level diagrams for $Z\to A A A$.} \label{treezaaa}
\end{figure}

It is interesting to note that generally the whole complete set
of diagrams for the scattering amplitude $e^- e^+ \to Z A A$
should be included in a calculation. For example, as shown in
Fig. \ref{lightbr}, when $\mu_{12}$ is about $58$ GeV,
 the coupling $\lambda_h$ vanishes for $M_h=130$ GeV, and
the bulk of the cross section comes from the diagrams (b) and (c)
of Fig. \ref{diagffzaa}. Furthermore, the effect of interference
among the complete set of diagrams can be so large that the
distribution of the production cross section as a function of
$\mu_{12}$ does not have a similar dip located the at the same
value of $\mu_{12}$. For example, as shown in
Fig.~\ref{sigmazaa05}, a broad dip in the distribution of
$\sigma(e^+e^- \to ZAA)$, predicted for LEP-2 with $M_h=130$ GeV
(the dashed curve), is located at $\mu_{12} \sim 42$ GeV, not
58 GeV.

As noted previously, the experimental signature of the $e^- e^+
\to Z A A$ event is the associated production of a $Z$ boson with
two energetic photons. Based on this class of data sample,
 the LEP-2 ``fermiophobic Higgs'' search has
imposed an upper limit on the decay branching ratio ${\mathrm
{Br}}(h \to \gamma \gamma)$ for a given $M_h$ \cite{lepfermiop}.
In that analysis, it was assumed that the production rate of $e^+
e^- \to Z h$ is the same as the SM and the decay width of $h$ is
identical to that of the SM. One can express the experimental
result on the photonic Higgs search in terms of the upper limit on
the product $\sigma(e^+ e^- \to Z h) \times {\mathrm {Br}}(h \to
AA \to "\gamma \gamma")$ as a function of $M_h$. This upper limit
can then constrain the allowed range of the parameter $\mu_{12}$
for a given $M_h$, which is shown in Fig. \ref{hmumh}. It turns
out that the LEP-2 ``fermiophobic Higgs'' search data is not
useful for constraining this model with $\tan\beta=2$. This is
because in that case the decay branching ratio of $A \to \gamma
\gamma$ is about 0.1 (cf. Fig. \ref{abran}) for $M_A=0.2$ GeV,
which largely reduces the $Z + "\gamma \gamma"$ event rate.
Nevertheless, for $\tan\beta=0.5$, only a small region of
$\mu_{12}$ is allowed for $M_h$ less than about
103 GeV.\footnote{ Based on the data presented in
Ref.~\cite{lepfermiop}, this is the highest value of $M_h$ for
which we can obtain a useful bound on the value of $\mu_{12}$. }

There is another important piece of data from LEP-2, that is the
LEP SM Higgs boson search based on $e^+ e^- \to Z \to Z h$ with
$h$ decaying into a $b \bar b$ pair. It was reported \cite{lepsmh}
that a handful events have been found to be compatible with the SM
Higgs cross section for $M_h$ about 115 GeV. Can a THDM with a
light $A$ be compatible with such an interpretation of the data?
One trivial answer is to have $M_h=115$ GeV with a choice of the
free parameter $\mu_{12}$ such that the decay branching ratio
${\mathrm {Br}}(h \to b \bar b)$ is about the same as that in the
SM. This would obviously require the range of $\mu_{12}$ to be
near the dip in Fig. \ref{lightbr}. For $\alpha=\beta-\pi/2$,
$\mu_{12}$ is about $53$ GeV for a 115 GeV light CP-even Higgs
boson $h$. This result implies a very specific production rate of
the diphoton pair produced via $g g \to h \to A A $ at the
Tevatron and the LHC. We shall come back to this production
process in the following sections.

Another solution to this question is to realize that the observed
$e^+ e^- \to Z h (\to b {\bar b})$ event rate at LEP-2 is
determined by the product $\sigma(e^+ e^- \to Z h) \times {\mathrm
{Br}}(h \to b \bar b)$. It can be the case that $M_h$ is less than
115 GeV so that $\sigma(e^+ e^- \to Z h)$ is larger than that for
the 115 GeV case. Because in a light $A$ THDM the additional
decay channel $h \to A A$ is available, the decay branching ratio
${\mathrm {Br}}(h \to b \bar b)$ decreases. This reduction can
compensate the increase in the production rate of $Zh$ to describe
the same experimental data.

In Fig. \ref{hmumh}, we show the corresponding range of the
parameters $\mu_{12}$ and $M_h$, assuming that the kinematic
acceptances of the signal and the background events do not change
largely as $M_h$ varies.\footnote{ While this assumption is valid
when $M_h$ is around 115 GeV, it is likely to fail when $M_h$ is
close to the $Z$-boson mass ($m_Z$). } Hence, if we follow the
LEP-2 conclusion on the SM Higgs boson search, i.e. at the 95\% CL
the current lower bound on $M_h$ is about 113.5 GeV, then the
product of $\sigma(e^+ e^- \to Z h) \times {\mathrm {Br}}(h \to b
\bar b)$ cannot be larger than the SM prediction for
$M_h=113.5$ GeV. For a given $M_h$, ${\mathrm {Br}}(h \to b \bar
b)$ cannot be too large. Therefore, this data could exclude the
values of $\mu_{12}$ near the dips shown in Fig. \ref{lightbr}. As
expected, this set of data and that for photonic Higgs search
provide a complementary information on constraining the model.
Combining these two sets of data, a light $h$ with mass less than
103 GeV in the type-II THDM is excluded when $\tan\beta=0.5$. For
$M_h > 103$ GeV, some constraints on the range of $\mu_{12}$ can
be obtained. The combined constraint is shown in Fig. \ref{hmumh}
for $\tan\beta=0.5$. For $\tan\beta=1$, a similar constraint can
be obtained, and the region with $M_h < 95$ GeV is excluded.

In the following analysis, we shall focus on the region of the
parameter space that is consistent with the LEP Higgs boson search
result, i.e. at the 95\% CL the current lower mass bound on a SM
Higgs boson is about 113.5 GeV, as discussed above.

\subsection{The Production Rate of $Z A A$ and $W A A$ at Tevatron and LHC}

At the Run-2 of the Fermilab Tevatron, a 2 TeV proton-antiproton
collider, the production rate of $p {\bar p} \to Z A A$ is about
0.1 pb for $M_h=110$ GeV, $\mu_{12}=100$ GeV, $M_H=1$ TeV,
and $\tan \beta=0.5$ or 2. Being a hadron collider,
Tevatron can also produce a light $A$ pair via the constituent
process $q' {\bar q} \to W^\pm A A$. The Feynman diagrams for this
scattering process are the same as those depicted in
Fig.~\ref{diagffzaa} after replacing $Z$ by $W^\pm$ everywhere and
$H$ by $H^\pm$ in diagram (b). For the same parameters given
above, the $W^\pm AA$ production rates is about 0.2 pb.
For completeness, we show in Figs. \ref{zaatev} and \ref{waatev}
the cross section for $Z AA$ and $W^\pm AA$ productions,
respectively, at the
Run-2 of the Tevatron, and the LHC (a 14 TeV proton-proton
collider). It is interesting to note that when the mass of the
charged Higgs boson is not too large (consistent with the
case of $\tan\beta=2$), and the coupling of
$h$-$A$-$A$ vanishes, the production rate of $W^\pm AA$ is
dominated by the associate production of $A$ and $H^\pm$ which
subsequently decays into a $W^\pm$-boson and $A$. Since the
experimental signal of $A$ is an ``isolated photon'', this signal
event appears as an event with a $W^\pm$ and two photons, hence,
its SM background rate is expected to be small.
For completeness, the squared amplitude for this partonic process is
also presented in Appendix~\ref{ampeezaa}.

\subsection{Other Production Mechanisms of a Light $A$ at Colliders}

In addition to the above production processes, a light $A$ can
also be copiously produced at hadron colliders, such as Tevatron
and LHC, via $q {\bar q},gg \to b {\bar b} A \, {\rm or} \, t
{\bar t} A$, $gg \to h  \to AA$, $gg \to H \to AA$, and $q {\bar
q} \to H^+ H^- \to W^+ W^- AA$. Because of the potentially large
background, the $b {\bar b} A$ mode is not likely to be
observable. However, the $t {\bar t} A$ mode can be easily
identified by requiring an isolated photon with a large transverse
momentum. At the Tevatron Run-2, the inclusive rate of $t {\bar t}
A$ with a 175 GeV top quark is 9.8(0.6) fb for $\tan
\beta=0.5(2)$, and at the LHC, it is 1.6(0.1) pb. The production
rates for the last two processes can be easily calculated by
multiplying the known cross sections for the production of $gg \to
h \,{\rm or} \, H\,$ \cite{ggh} and $q {\bar q} \to H^+ H^-\,$
\cite{hphm} by the relevant decay branching ratios (given in the
previous sections). For example, at the Tevatron Run-2, the
production cross section of $gg \to h \to AA$ is $0.8$ and $0.5$
pb for $M_h=110$ GeV and 130 GeV, respectively, with
$\mu_{12}=100$ GeV and $\tan \beta=0.5$ or 2.
Since we have set $\alpha=\beta-\pi/2$
(motivated by the $\rho$-parameter constraint) in all our
calculations, the production rate of $gg \to h$ is independent of
$\tan\beta$. Furthermore, when $\mu_{12}=100$ GeV, the decay
branching ratio for $h \to AA$ is about 1, cf. Fig.~\ref{lightbr}.
Therefore, the above rates for $\tan \beta=0.5$ or 2 are about the
same. At the LHC, the rates are $35$ pb and $27$ pb for
$M_h=110$ GeV and 130 GeV, respectively. Hence, this mode is
important for further testing the THDM with a very light $A$. In
Fig.~\ref{gghaa}, we show the production rate of $gg \to h \to AA$
 at the Tevatron Run-2 and the LHC as a
function of $\mu_{12}$ for various $M_h$.\footnote{
To compare with the experimental data, one
should also include the decay branching ratio ${\mathrm {Br}}(A
\to \gamma \gamma)$, cf. Fig. \ref{abran},
 for each CP-odd scalar
$A$ decaying into its photon mode.}

The signal rate of $gg \to H \to AA$ is different from that of $gg
\to h \to AA$ because the decay branching ratio of $H \to AA$ is
not the same as that of $h \to AA$ and the coupling of $H$ to $t$
in the loop has a factor of $\cot \beta$, cf.
Eq.~(\ref{eq:ffs_coup}). For instance, the branching ratio of $H
\to AA$ is about $0.12$ ($0.47$) for $M_H=180$ GeV ($350$ GeV)
with $\mu_{12}=100$ GeV, so that the cross section of $gg \to H
\to AA$ is about $6$ fb ($2.6$ fb) at the Tevatron, and
$0.48$ pb ($0.89$ pb) at the LHC when $\tan\beta=2$. For
$\tan\beta=0.5$, the low energy data requires $M_H$ to be around
1 TeV, so that its rate is negligible.
 In Fig.~\ref{gghhaa}, we show the
$p{\bar p},pp(gg) \to H \to AA$ production cross section as a
function of $\mu_{12}$ at the Tevatron Run-2 and the LHC for a few
values of $M_H$, with $M_A=0.2$ GeV and $\tan \beta =2$.
Note that when the rate of $gg \to H \to AA$ is small (for certain
values of $\mu_{12}$), the rate of $gg \to H \to AZ$ becomes large
because the sum of the decay branching ratios of these two modes
is about 1, cf. Fig.~\ref{heavybran2}.
Hence, their roles to the discovery of a heavy Higgs boson
in this model are complementary to each other.

Usually, a charged Higgs boson $H^+$ in the THDM is assumed to
decay via the heavy fermion pairs, either the $t \bar b$, $c \bar
b$ or $\tau^+ \nu$ modes. However, as shown in Fig.~\ref{hcbraii},
when $A$ is light, the decay mode of $W^+ A$ can become dominant.
In that case, the scattering process $q {\bar q} \to H^+ H^- \to
W^+ W^- AA$ will be seen by the detector as a $W$-boson pair with
two isolated photons. With a proper kinematic cut, this event can
be separated from its SM backgrounds. (Although we have limited
ourselves to the discussion on the decay mode of $A \to \gamma
\gamma$, the other decay mode into an $e^+e^-$ pair can also prove
to be useful for testing such a model.) For $M_{H^+}=180$ GeV
($350$ GeV), the decay branching ratio of $H^- \to W^- A$ is
about 1 (0.8), so that the cross section of $q {\bar q} \to H^+
H^- \to W^+ W^- AA$ is about $1.7$ fb ($0.01$ fb) at the
Tevatron, and $31$ fb ($1.5$ fb) at the LHC when $\tan\beta=2$.
Again, when $\tan\beta=0.5$, the low energy data requires
$M_{H^+}$ to be around 1 TeV, so that its rate is negligible.

In the THDM, there is no tree level coupling $Z$-$W^\pm$-$H^\mp$,
therefore the cross section for the scattering process $q {\bar q}
\to W^\pm H^\mp \to W^+ W^- A$ at the tree level is dominated by
the bottom quark fusion $b {\bar b} \to W^\pm H^\mp \to W^+ W^-
A$. At the Tevatron, its rate is negligible (about 0.1 fb), and
at the LHC, its rate is about 0.1 pb for $M_{H^+}=200$ GeV, and
$\tan\beta=2$.

At the future Linear Collider (a 500 GeV $e^+e^-$ collider), the
$ZAA$ production rate is shown in Figs.~\ref{sigmazaa} and
\ref{sigmazaa05} for a few choices of parameters. Besides this
production mode, a light $A$ can also be copiously produced via
$e^+ e^-  \to H^+ H^- \to W^+ W^- AA$ provided that the cross
section $e^+ e^-  \to H^+ H^-$ is not small. For
$M_{H^+}=180$ GeV (200 GeV), we expect that rate of $W^+ W^- AA$
to be about 41 fb (22 fb) for $\tan\beta=2$. (The Br($H^+ \to
AW^+$) is about 1 and 0.9 for $M_{H^+}=180$ GeV and 200 GeV,
respectively.)

\section{Discussion and Conclusion}
\label{conclu}
\indent

In this paper, we examined the possibility of having a very light
CP-odd scalar $A$ (with a mass about 0.2 GeV)
in a general THDM. After examining the relevant low energy data,
we found that this model is either excluded already (according to
the DH prediction of the SM contribution to the muon $(g-2)_\mu$) or
its parameter space has been largely constrained.
Assuming the CLY prediction of the SM contribution,
the muon $(g-2)_\mu$ data requires $\tan \beta < 2.6$, regardless
of the other parameters of the type-II THDM. (For the type-I THDM,
the muon $(g-2)_\mu$ data requires $\tan\beta > 0.4 $.)
For such a light $A$, the CUSB data on Br($\Upsilon \to A \gamma$)
requires $\tan\beta$ to be around 1. For a type-II THDM, the $b \to
s \gamma$ data requires $M_{H^+}$ to be larger than about 165 GeV
to 200 GeV when $\tan\beta$ is larger than 1. For a type-I THDM,
the constraint on $M_{H^+}$ is much looser. The $\rho$-parameter
data also imposed a stringent constraint on the difference between
$M_{H}$ and $M_{H^+}$ for a given value of $\sin(\beta-\alpha)$.
For $\sin(\beta-\alpha)=1$, $M_{H}$ and $M_{H^+}$ have to be almost
equal. The $R_b$ data also provides a stringent bound on the model.
For example, when $\tan\beta=0.5$, the $R_b$ data requires $M_{H^+}$
to be around 1 TeV in the THDM. Consequently, due to the
$\rho$-parameter constraint, $M_{H}$ should also be around 1 TeV.
A summary of the low energy constraints
on this model is given in Figs.~\ref{gm2tbma} -- \ref{rbounds}
as well as Table~\ref{allconstraints}.

After finding the allowed parameter space of the model, we examine
the impact on the decay branching ratios and total decay width of
the Higgs bosons due to the presence of a light $A$. Depending on
the value of the soft-breaking
parameter $\mu_{12}$, present in the Higgs
potential of a general THDM, the total decay width of $h$, $H$ or
$H^+$ can become large because of the large phase space volume for
the decay channels
($h \to AA, AZ$), ($H \to AA, AZ$), or ($H^+ \to A W^+$). To have a
valid perturbative calculation, we only consider the values of
$\mu_{12}$ such that the total decay width of the Higgs bosons is
small as compared to its mass. Due to the small mass of $A$, the
decay branching ratios for the decay modes $AA$, $AZ$, or $AW^\pm$
can be sizable, which can result in a very different detection
mode for the THDM. In Figs.~\ref{lightbr}-\ref{heavybran2}, we
showed a few of such examples.

The exciting feature of such a light $A$ is that when
it decays into a photon pair, because of the typical large energy
of $A$ produced from the decay of other heavy Higgs bosons, its
decay photon pair will register in the detectors as a single
photon signature. Hence, the SM background rate for detecting such
a signal event is expected to be generally small. Therefore, the
Tevatron Run-2, the LHC and the future LC have a great potential
to either detect a light $A$ in the THDM or to exclude such a
theory model. A few potential discovery modes at various colliders
were given in Section IV, cf. Figs.~\ref{sigmazaa}-\ref{gghhaa}.

%%%%%%%%%%%%%%%%%%%%%%%%%%%%%%%%%%%%%%%%%%%%%%%%%%
\section*{ \bf  Acknowledgments }
\indent \indent  FL and GTV would like to thank CONACYT and SNI (M\'exico)
for support. The work of CPY was supported in part by NSF grant
PHY-9802564. FL and CPY thank CERN for hospitality, where part of
this work was completed. CPY also thank the warm hospitality of
the National Center for Theoretical Sciences in Taiwan.

\vspace*{3.5mm} \noindent \underline{Note Added:}
\linebreak\hspace*{2.5mm}

During the preparation of this manuscript, two similar papers
\cite{add} were posted to the hep-ph archive very recently. For
the part we overlap, our results agree. Furthermore, after the
submission of this paper, a new article \cite{gambino} concluded a
tighter bound on $M_{H^+}$ from $b \to s \gamma$ data than the one
\cite{borzumati} adopted in our analysis. Nevertheless, the
general conclusion about the phenomenology of a light $A$
discussed in this paper remains unchanged.

%%%%%%%%%%%%%%%%%%%%%%%%%%%%%%%%%%%%%%%%%%%%%%%%%

\appendix

\renewcommand{\theequation}{\Alph{section}.\arabic{equation}}

\section{Anomalous magnetic moment of muon in the THDM}
\label{magmoment}

Given the THDM interaction Lagrangian

\begin{equation}
{\cal L}=\frac{-g m_\mu}{2 m_W}\left(\sum_{S=H,h} C_S \bar\mu \mu
S + i\,C_A \bar\mu\gamma^5\mu A+\left(\sqrt{2}\, C_{H^\pm}\bar\mu
\nu_{L} H^{-}+\mathrm{h.c.}\right)\right),
\end{equation}

\noindent the contribution from the diagrams depicted in
Figs. \ref{magmom} (a) and (b) reads
\cite{oneloop}

\begin{equation}
a_\mu^{\mathrm one-loop}=\frac{\alpha\,m_\mu^2}{8 s_w^2 m_W^2 \pi}
\, \sum_{j} \int^1_0  \, dx \,C^2_j\,g^j(x),
\end{equation}

\noindent where $j$ is summed over $h$, $H$, $A$, and $H^\pm$. The
respective $g^j(x)$ are:

%\begin{mathletters}
\begin{equation}
g^{H,h}(x)= \frac{x^2\left(2-x\right)}{x^2+\Lambda_{H,h} (1-x)},
\end{equation}

\begin{equation}
g^{A}(x)= \frac{-x^3}{x^2+\Lambda_A (1-x)},
\end{equation}

\begin{equation}
g^{H^\pm}(x)= \frac{x(x-1)}{x+\Lambda_{H^\pm}-1},
\end{equation}
%\end{mathletters}

\noindent with $\Lambda_j=(M_{j}/m_\mu)^2$. The respective $C_j$ are
given in Table \ref{couplings} for a type-I or
type-II THDM.

As for the contribution from the Barr-Zee diagram depicted in
Fig.~\ref{magmom} (c), the dominant two-loop contribution
comes from a CP-odd scalar in the loop.
There is another diagram with the virtual photon replaced
by a $Z$ gauge boson. Because it is highly suppressed, we will
not take it into account.
According to \cite{darwin}, the dominant two-loop contribution
to $a_\mu$ is given by

\begin{equation}
a_\mu^{\mathrm two-loop}=\frac{\alpha^2}{8 \pi^2
s_w^2}\frac{m_\mu^2 C_\mu}{M_W^2}\sum_{f=t,b,\tau} N_c^f Q_f^2 C_f
\chi_f f\left(\chi_f\right),
\end{equation}

\noindent with $\chi_f=(m_f/M_A)^2$, $N_c^f=1 \,(3)$ for leptons
(quarks), $m_f$ and $Q_f$ are the mass and charge of fermion, and
$C_f$ is given by the interaction Lagrangian of the CP-odd scalar to fermions:
\begin{equation}
{\cal L}=i\frac{g \, C_f \, m_f}{2\,m_W} \bar f \gamma^5 f A,
\end{equation}

\noindent where $C_t=\cot\beta$ for either a type-I or type-II
THDM, whereas $C_f=-\cot \beta$ and $\tan\beta$ for a type-I and
type-II THDM, respectively, for $f=b$, $\mu$, and $\tau$. Finally,
$f(x)$ is given by

\begin{equation}
f(x)=\int^1_0 \,dz \frac{\log\left(\frac{x}{z(1-z)}\right)}{x-z(1-x)}.
\end{equation}

For comparison, we note that if only one-loop contributions are
included in our analysis, we will derive a different constraint on
the parameters of $\tan\beta$ and $M_A$. In that case, the figures
corresponding to Figs.~\ref{gm2tbma-1} and \ref{gm2tbma-2} are
shown as Figs. ~\ref{gm2tbma-one-1} and \ref{gm2tbma-one-2}.

%%%%%%%%%%%%%%%%%%%%%%%%%%%%%%%%%%%%%%%%%%%%%%%%%

\section{The fermion-loop contribution to the $Z\to AAA$ decay}
\label{zaaabox}

%%%%%%%%%%%%%%%%%%%%%%%%%%%%%%%%%%%%%%%%%%%%%%%%%%%%%%%%

Under the scenario that
the branching ratio of $A \to \gamma
\gamma$ is close to unity, it is possible that the CP-odd scalar
$A$ has a large coupling to fermions only. This can happen
when the  other scalars do not exist at all or are so heavy that
they would decouple from the low energy effective theory because of
their small couplings to $A$. We consider this assumption to examine the
contribution of fermion loops to the decay
$Z \to A A A$.\footnote{The decay mode $Z \to A A$ is forbidden
by the Yang-Landau theorem.}
One of the fermion loop diagrams contributing to the $Z(p) \to
A(k_1)A(k_2)A(k_3)$ process is shown in Fig.~\ref{diag}. (There are
five other diagrams with the obvious permutations of the
pseudoscalar boson momenta.)

The fermion loop contribution to the decay of
$Z \to A A A$ was first roughly estimated in \cite{li},
and then was re-examined in \cite{chang} by considering
 only the top quark loop contribution.
A complete calculation, including also bottom quark contribution
with a large $b\bar{b}A$ coupling, was never presented in the
literature. We consider an effective theory in which the coupling
of the CP-odd scalar to quarks is $C_q m_q/v$, where $v(=2 m_W/g)$
is the vacuum expectation value, $m_q$ is the mass of the quark,
and the coefficient $C_q$ depends on the choice of models. For a
SM-like coupling, assuming the existence of $A$, $C_q=1$ for both
up- and down-type quarks.
For a type-I or type-II THDM, $C_q=\cot\beta$ for
up-type quarks. For down-type quarks,
$C_q=-\cot \beta$ and $\tan\beta$ for a type-I and type-II
THDM, respectively,
Since the coupling of $Z$ to up- and down-type quarks
is fixed by the $ \mathrm{SU(2)_{L} \times U(1)_{Y}} $ gauge
interaction, the effective Lagrangian can be written as

\begin{equation}
{\cal L}=\sum_{q} \left[  \frac{g}{2 c_w} \bar{u}_q \gamma^\alpha
(g_V^q-g_A^q \gamma^5) u_q \ Z_\alpha + \frac{i g \ m_q C_q}{2
m_W} \bar{u}_q  \gamma^5 u_q \ A\right].
\end{equation}
Given the above interaction, the $Z \to AAA$ amplitude can be
expressed, in terms of Mandelstam variables $s=(k_1+k_2)^2$,
$t=(k_2+k_3)^2$, and $u=(k_1+k_3)^2$, as

\begin{equation}
\label{amptot} {\cal{M}}= \frac{3 \alpha^2 m_Z}{16 s_w^4 c_w^4}
\Big[F(t,u) k_1^\alpha+F(u,t) k_2^\alpha \Big] \epsilon_\alpha
(p),
\end{equation}

\noindent where

\begin{eqnarray}
\label{facform} F(t,u)&=&\sum_q C_q^3 \left(
\frac{m_q}{m_Z}\right)^4 \Big[ (4M_A^2-s-t) \big(D_0(s,u)-
D_0(t,u)\big)+
(t-s)D_0(s,t) \nonumber \\
&+& 4\big(C_0(t)- C_0(s)\big) \Big],
\end{eqnarray}

\noindent with the scalar three- and four-point integrals given by

\begin{equation}
C(s)=\frac{1}{i \pi^2}\int \frac{d^4k}{\left[k \right]
\left[k-k_3\right]\left[k-p\right]},
\end{equation}

\begin{equation}
D_0(s,t)=\frac{1}{i \pi^2}\int \frac{d^4k}{\left[k\right]
\left[k-k_1\right] \left[k-k_1-k_2\right]\left[k-p\right]},
\end{equation}

\noindent where we use the short-hand notation
$[\ell]=\ell^2-m_q^2+i\epsilon$. The remaining scalar integrals
can be obtained by permuting the pseudoscalar bosons momenta. The
squared amplitude, after averaging over the $Z$ boson
polarizations, is
\begin{eqnarray}
\label{decay} \frac{1}{3}\sum_{spins} |{\cal M}|^2 &=& 3 \left(
\frac{\alpha^2}{16 s_w^4 c_w^4} \right)^2 \Big[
((m_Z-M_A)^2-t)((m_Z+M_A)^2-t)|F(t,u)|^2 \nonumber \\
&+& \left((m_Z^2-M_A^2)(t+u)+M_A^4+tu \right)
{\mathrm{Re}}{\left(F(t,u)F^{\dag}(u,t) \right)} +\left(t
\leftrightarrow u \right)\Big].
\end{eqnarray}

Given this result, the decay width can be computed by the usual
methods. Although our result is quite general, we only expect
important contributions from both $b$ and $t$ quark loops:
contributions from lighter quarks are suppressed by the
factor $(m_q/m_Z)^4$ and can be ignored unless $C_q$ is extremely
large, which however is unlikely to be true because of the
tree-level unitarity constraint on the couplings. As we are
primarily interested in examining the case of a very light
pseudoscalar Higgs boson, we can safely neglect $M_A $ in our
calculation. In the $M_A \to 0$ limit the scalar integrals
become \cite{van}

\begin{equation}
C_0(s)=\frac{-2}{s-m^2_Z}\left[F\left(\frac{4 m^2_q}{s}\right)
-F\left(\frac{4 m^2_q}{m^2_Z}\right)\right],
\end{equation}

\begin{equation}
D_0(s,t)=\frac{2}{s t}\left[G(s,t,s)+
G(s,t,t)-G(s,t,m_Z^2)\right],
\end{equation}

\noindent where $F$ is given in Eq. (\ref{Ffun}), and $G$ can be
written in terms of Spence functions

\begin{eqnarray}
G(s,t,y)&=&\frac{1}{\phi}
\left\{{\mathrm{Sp}}\Big(\frac{a_-}{a_--b} \Big) -
{\mathrm{Sp}}\Big(\frac{a_+}{a_+-b} \Big)
+{\mathrm{Sp}}\Big(\frac{a_-}{b-a_+} \Big)
\right. \nonumber \\
&-& \left. {\mathrm{Sp}}\Big(\frac{a_+}{b-a_-} \Big)+
\log\Big(\frac{-a_-}{a_+} \Big) \log
\Big(1-i\epsilon-\frac{y}{m_f^2} a_-a_+ \Big) \right\},
\end{eqnarray}

\noindent with $\phi=\sqrt{1+4 m^2_q \ (m_Z^2-s-t) /(s \ t)}$,
$a_\pm=\frac{1}{2}\big(1 \pm \phi \big)$/2, and
$b=\frac{1}{2}\big(1 + \sqrt{1-4(m^2_q-i\epsilon)/y} \big)$.

It is easy to estimate the order of the scalar integrals arising
from a heavy quark: if $m_q > 2 m_Z$ and $M_A \to 0$, we can use
the heavy mass expansion approximation. The leading term of the
three-point scalar integral is $C(s,t) \sim -1/(2 m^2_q) $, which
can be differentiated with respect to $m_q^2$ to give $D(s,t) \sim
1/(6m^4_q )$. We then have $C(s) \sim -1.62 \times 10^{-5}$
GeV$^{-2}$ and $D(s,t)\sim 1.78 \times 10^{-10}$ GeV$^{-4}$ for
$m_q=175$ GeV, the top quark case. On the other hand, for the $b$
quark loop, numerical evaluation gives $|C(s)| \sim
10^{-4}-10^{-3}$ GeV$^{-2}$ and $|D(s,t)| \sim 10^{-6}-10^{-5}$
GeV$^{-4}$, which indicates that bottom quark contributions may
compete with that of the top quark, even if we consider the factor
of $(m_q/m_Z)^4$ in (\ref{facform}). In conclusion, the decay
branching ratio due to the top and bottom quark loop contributions
is
\begin{equation}
\label{br} {\mathrm {Br}}(Z \to AAA)=1.3 \times 10^{-18}
C_t^6+2.47 \times 10^{-17} C_b^6 + 7.63 \times 10^{-18} C_t^3
C_b^3
\end{equation}

\noindent which is many orders below the estimate given before.
However, there is no contradiction because the authors of
\cite{chang} used a rough estimate to show that top quark loops
cannot enhance the branching fraction of $Z \to AAA$ beyond
$10^{-10}$ in the case of $C_t=1$. From (\ref{br}) we can see that
top quark contribution is smaller unless $C_t \gg 1$, while bottom
quark contribution is larger for $C_b > C_t$. A large $C_b$ in a
type-II THDM-like model implies large $\tan \beta$; however,
because of the unitarity bound,
the coupling $C_b$ cannot be arbitrary large.
By requiring the validity of a perturbation calculation, we can derive
the upper bound on $C_b$ to be about 120 which yields
${\mathrm Br}(Z \to AAA)\sim 10^{-5}$.

\section{The processes $\lowercase{{\bar f}f} \to ZAA$ and
$\lowercase{{\bar q}q^{\prime}}
\to W^+AA$}
\label{ampeezaa}

\subsection{Squared Amplitude of ${\lowercase{{\bar f}f}} \to Z A A$}

The Feynman diagrams contributing to the scattering of
${\bar f}f \to Z A A$ are shown in Fig.~\ref{diagffzaa}.
The scattering amplitude for
${\bar f}(p_1) f(p_2) \to Z(k) A(k_1) A(k_2)$
can be written as:
\begin{equation}
\label{amp} {\cal M}= \bar{u}_f  \gamma_{\mu} \left(g^f_L
P_L+g^f_R P_R \right) u_f \, \Big[ F_0 \, g^{\mu \nu} + F_1 \,
k_2^{\mu} k_1^{\nu} +F_2 \, k_1^{\mu} k_2^{\nu}\Big]
\epsilon^*_{\nu}(k),
\end{equation}
\noindent where $g_L^f=I_3-e_f s_w^2$ and $g_R^f=-e_f s_w^2$, with
$I_3$ the weak isospin of the fermion and $e_f$ its electric
charge in units of positron's. After averaging over the spins and colors
of the initial state and summing over the polarizations of the final
state particles, we obtain

\begin{equation}
\label{sqamp} \frac{1}{4}\sum_{spins}|{\cal M}|^2=\frac{N_c g^6 \,
 \, \chi^2 } {4  \,m_Z^2 \, (s-m_Z^2)^2}\left[ \sum_{i=0}^2
\zeta_i |\xi_i|^2+ \sum_{i,j=0, \, j>i}^2 \zeta_{ij} {\mathrm Re}
\left( \xi_i \xi_j \right) \right],
\end{equation}
\noindent where $N_c=1$ or $1/3$ for leptons or quarks,
respectively. Furthermore
\begin{equation}
\label{chi2} \chi^2=\frac{I_3^2-2 e_f I_3 s_w^2+ 2 e_f^2
s_w^4}{c_w^6},
\end{equation}

\noindent whereas the $\zeta_i$ are

\begin{mathletters}
\label{zetas}

\begin{equation}
\zeta_0=4 (k \cdot p_1) (k \cdot p_2) + m_Z^2 \, s,
\end{equation}

\begin{equation}
\zeta_1=\left(M_A^2 \, s - 4 (k_2 \cdot p_1)( k_2 \cdot p_2)
\right) \left(M_A^2 m_Z^2 - (k \cdot k_1)^2 \right),
\end{equation}

\begin{equation}
\zeta_2=\zeta_1(k_1 \leftrightarrow k_2),
\end{equation}

\begin{equation}
\zeta_{01}=2\left( m_Z^2 \, \eta(k_1) -(k \cdot k_1) \, \eta(k)
\right),
\end{equation}

\begin{equation}
\zeta_{02}=\zeta_{01}(k_1 \leftrightarrow k_2),
\end{equation}

\begin{equation}
\zeta_{12}=2 \left(m_Z^2 (k_1 \cdot k_2)- (k \cdot k_1)(k \cdot
k_2) \right) \eta(k_1),
\end{equation}

\begin{equation}
\eta(\ell)=s \,(\ell \cdot k_2) - 2\Big( (\ell \cdot p_1) (k_2
\cdot p_2)+ (\ell \cdot p_2) (k_2 \cdot p_1) \Big),
\end{equation}

\end{mathletters}

\noindent and $s=(p_1+p_2)^2$.
 The form factors $\xi_i$ are given by

\begin{equation}
\label{xi} \xi_i \equiv \xi_i^Z= \xi_i^h+\xi_i^H.
\end{equation}

\noindent Diagrams with the $ZZAA$ and $ZZH(h)$ vertices
contribute to the $\xi_0$ form factor, whereas those with $ZAH(h)$
vertex contributes to $\xi_{1,2}$

\begin{mathletters}
\label{xis}

\begin{equation}
\xi_0^{\phi}= \frac{1}{2}-\frac{1}{(s_2-M_{\phi}^2)}
\left\{\begin{array}{lr} \frac{s_{\delta_-}}{s^2_{2 \beta}}\left(
2 c_{\delta_+}\mu_{12}^2-s^2_{2 \beta} s_{\delta_-}M_A^2-s_{2
\beta} \left(c_{\alpha}c_{\beta}^3-s_{\alpha} s_
{\beta}^3\right)M_h^2
\right)& \phi=h, \\ \\
\frac{c_{\delta_-}}{s^2_{2 \beta}}\left( 2
s_{\delta_+}\mu_{12}^2-s^2_{2 \beta} c_{\delta_-}M_A^2-s_{2 \beta}
\left(c_{\alpha}s_{\beta}^3+s_{\alpha} c_ {\beta}^3\right)M_H^2
\right)& \phi=H,
\end{array}
\right. \label{xi0}
\end{equation}

\begin{equation}
\xi_{1,2}^{\phi}=\frac{1}{(k+k_{1,2})^2-M_{\phi}^2}
\left\{\begin{array}{lr}
c_{\delta_-}^2& \phi=h, \\ \\
s_{\delta_-}^2& \phi=H,
\end{array}
\right.
\end{equation}

\end{mathletters}

\noindent where $\delta_\pm=\beta \pm \alpha$ and
$s_2=(k_1+k_2)^2$. In addition, it is understood that the Higgs
propagators acquire an imaginary part in the resonance region,
{\it i.e.} $p^2-M_{\phi}^2 \to p^2-M_{\phi}^2 + i \Gamma_{\phi}
M_{\phi}$, where $\Gamma_{\phi}$ is the total width of $\phi$.

\subsection{Squared Amplitude of ${\bar q}q^{\prime} \to W^+AA$}

\begin{figure}
\centerline{\hbox{ \epsfig{figure=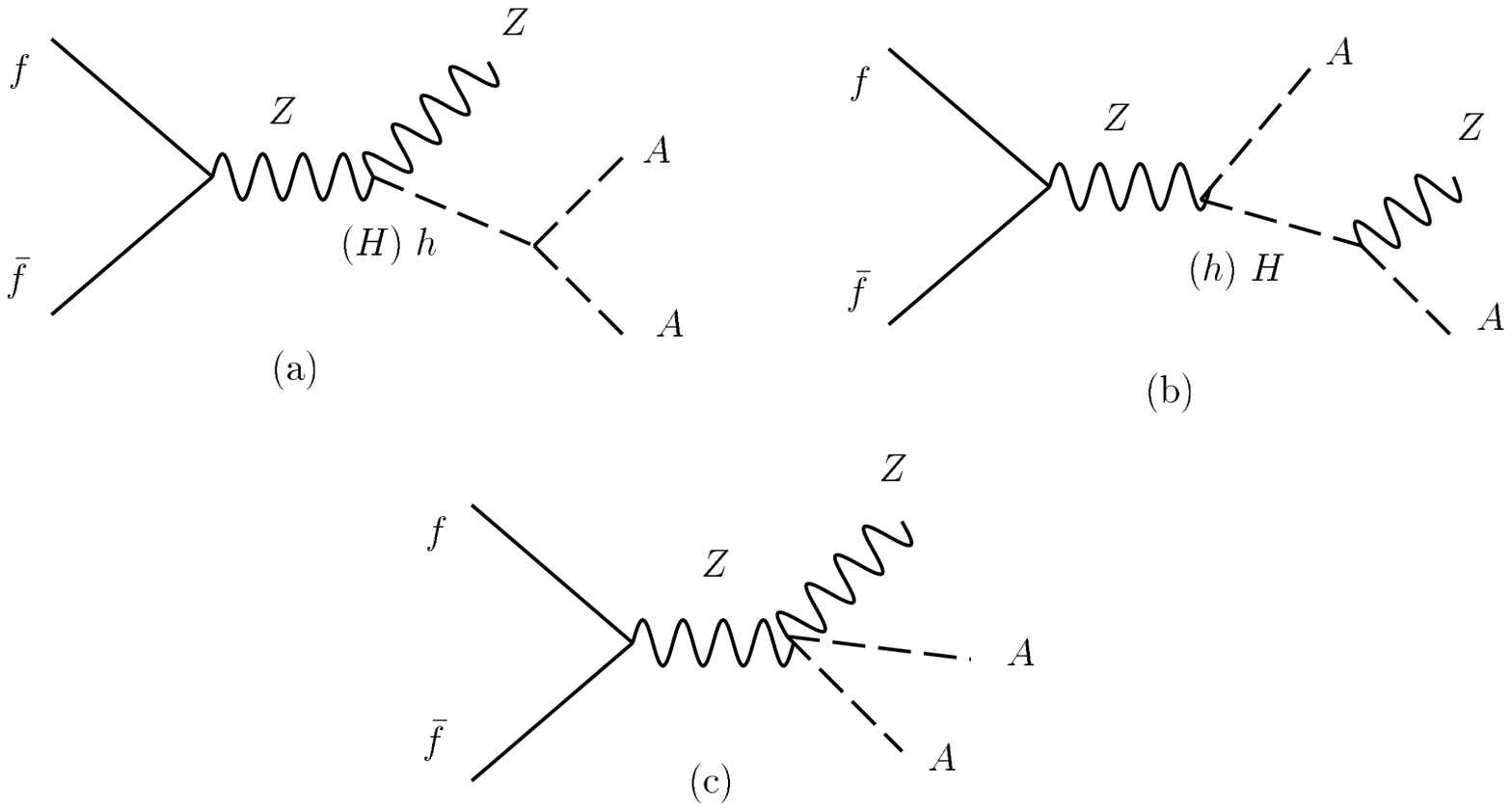,height=2.5in}}}
\caption{Diagrams for $f {\bar f} \to Z A A$.} \label{diagffzaa}
\end{figure}

\begin{figure}
\centerline{\hbox{ \epsfig{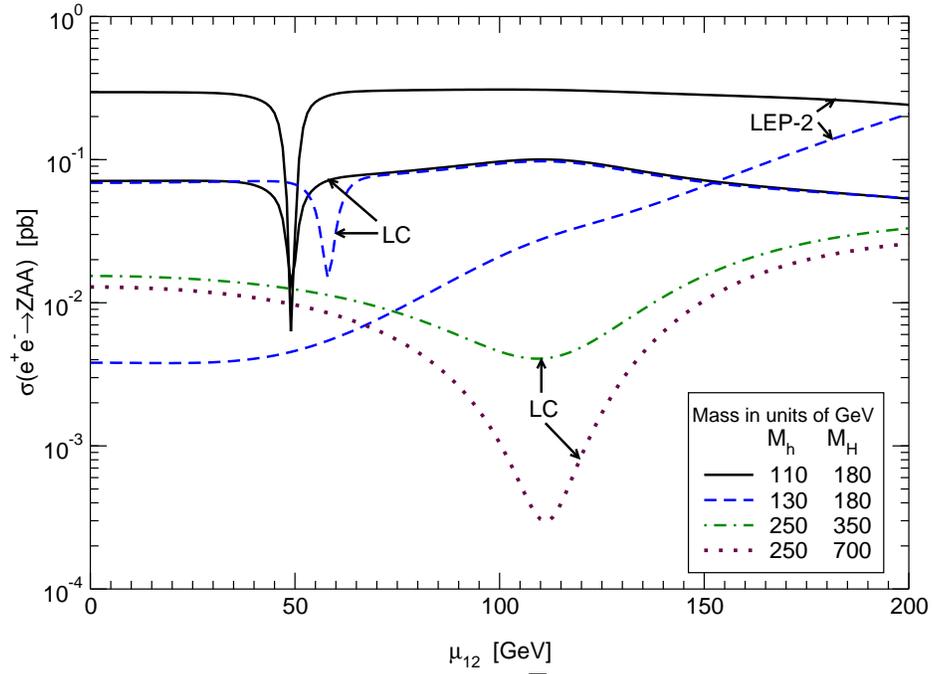}}}
\caption{ $ZAA$ production cross section at LEP-2 and LC with
$\sqrt{S}=210$ GeV and 500 GeV, respectively,
 for a few values of $M_h$ and $M_H$
 with $M_A=0.2$ GeV and $\tan \beta =2$. }
\label{sigmazaa}
\end{figure}

\begin{figure}
\centerline{\hbox{ \epsfig{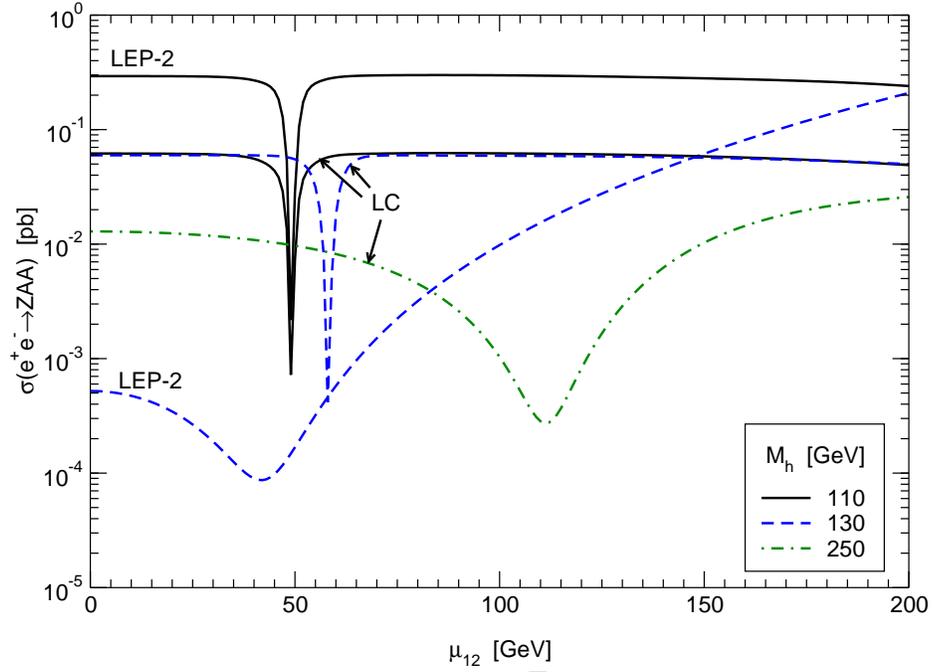}}}
\caption{ $ZAA$ production cross section at LEP-2 and LC with
$\sqrt{S}=210$ GeV and  500 GeV, respectively, for a few values
of $M_h$
 with $M_A=0.2$ GeV, $M_H=1$ TeV and $\tan \beta =0.5$. }
\label{sigmazaa05}
\end{figure}

\begin{figure}
\centerline{\hbox{ \epsfig{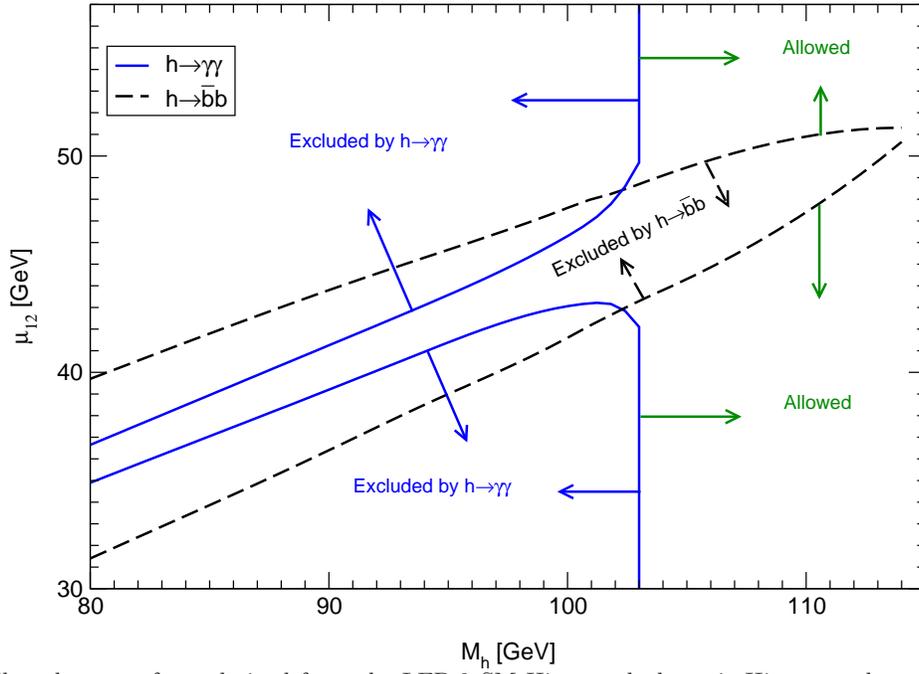}}}
\caption{ The allowed range of $\mu_{12}$ derived from the LEP-2
SM Higgs and photonic Higgs search results as a function of $M_h$,
with $M_A=0.2$ GeV and $\tan \beta=0.5$. For $M_h < 120$ GeV,
this range is not sensitive to the actual value of $M_H$. }
\label{hmumh}
\end{figure}

\begin{figure}
\centerline{\hbox{ \epsfig{figure=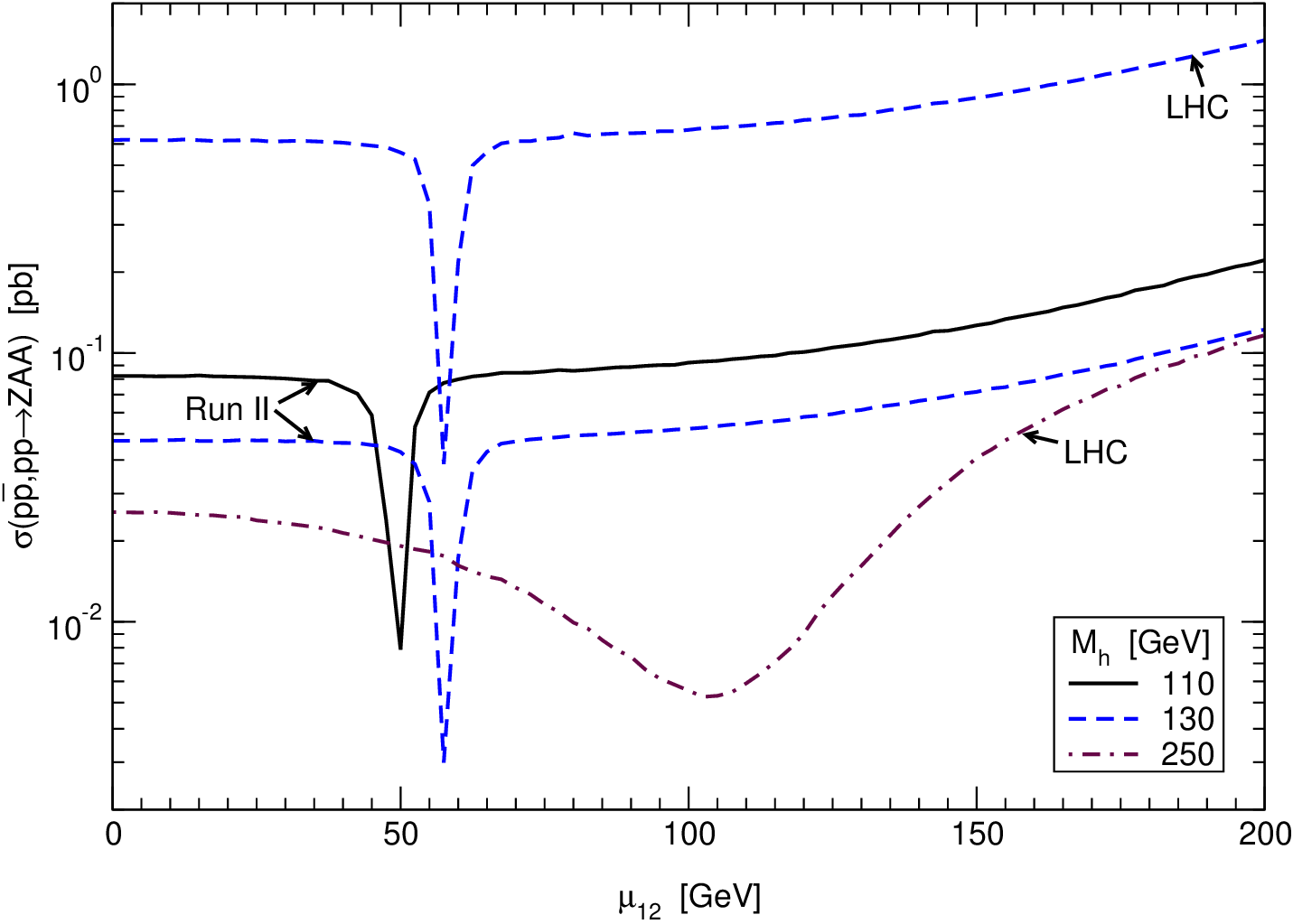,height=3.5in}}}
\caption{ $Z AA$ production cross section at the Run-2 of the
Tevatron and the LHC, for a few values of $M_h$, with
$M_{H}=1$ TeV, $M_A=0.2$ GeV and $\tan \beta =0.5$.}
\label{zaatev}
\end{figure}

\begin{figure}
\centerline{\hbox{ \epsfig{figure=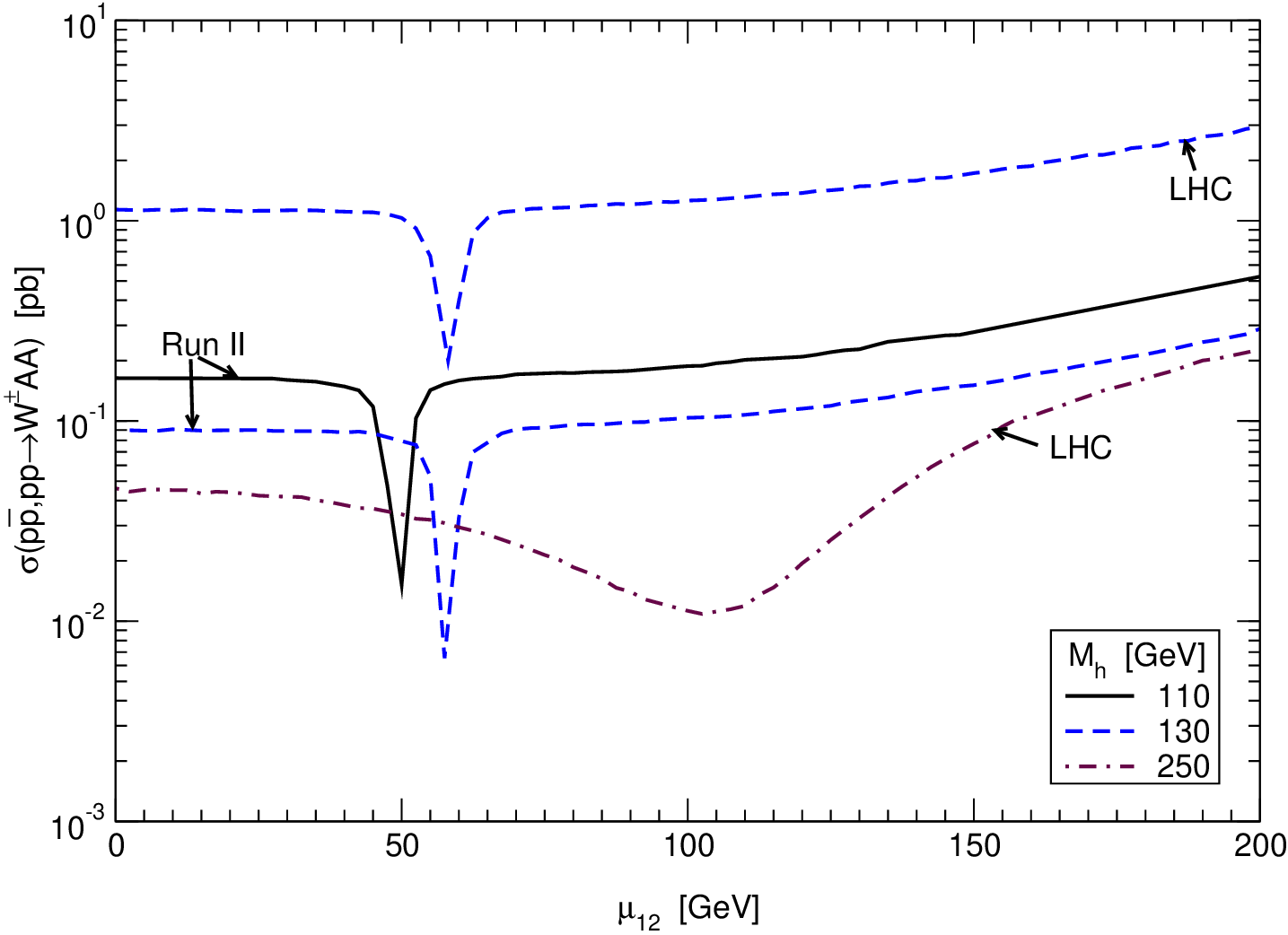,height=3.5in}}}
\caption{ $W^\pm AA$ production cross section at the Run-2 of the
Tevatron and the LHC, for a few values of $M_h$, with $M_{H^\pm}=
1$  TeV, $M_A=0.2$ GeV and $\tan \beta =0.5$.} \label{waatev}
\end{figure}

\begin{figure}
\centerline{\hbox{ \epsfig{figure=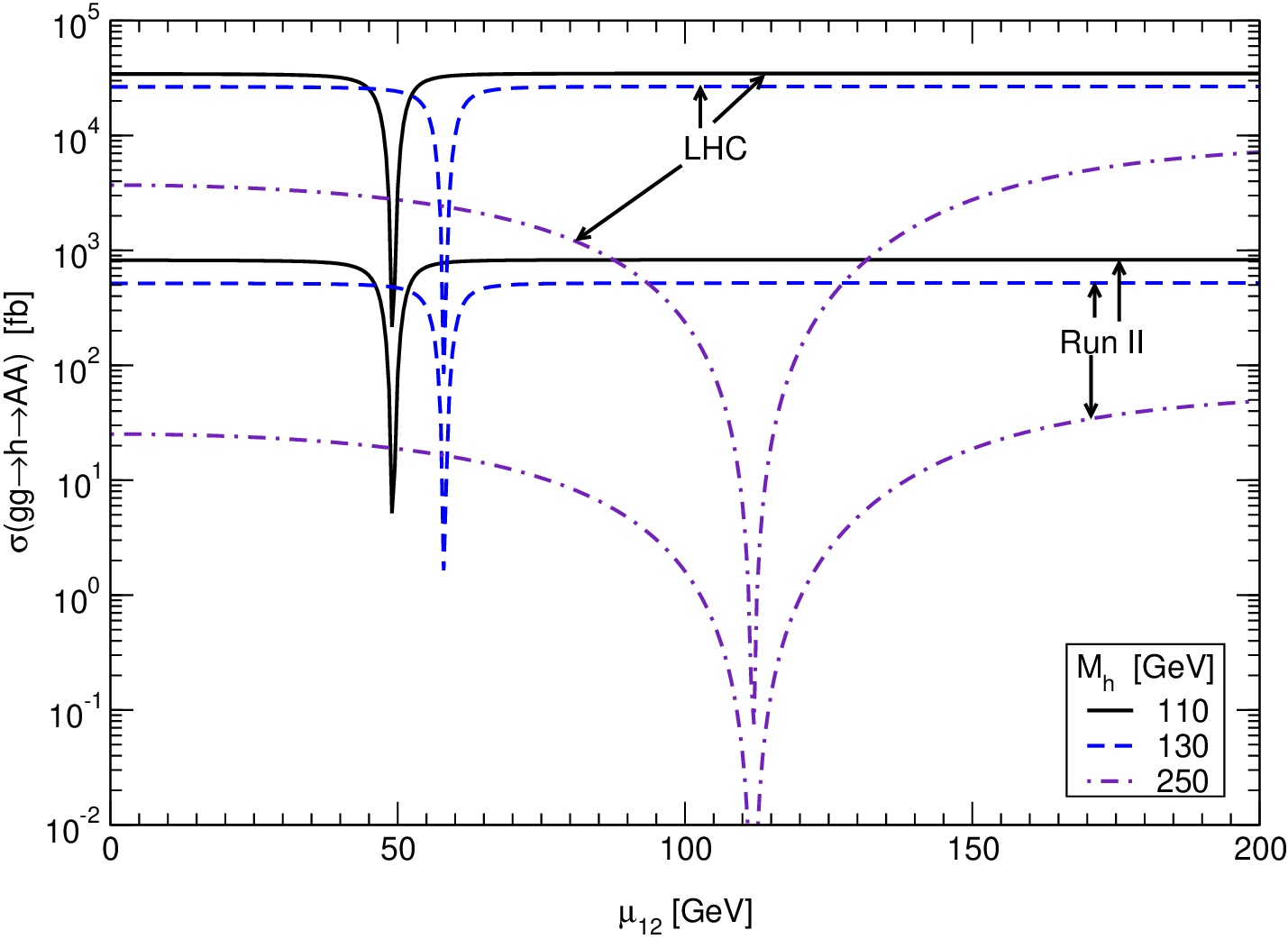,height=3.5in}}}
\caption{ The $p{\bar p},pp(gg) \to h \to AA$ production cross
section as a function of $\mu_{12}$ at the Tevatron Run-2 with
$\sqrt{S}=2$ TeV and the LHC with $\sqrt{S}=14$ TeV, for a few
values of $M_h$ with $M_A=0.2$ GeV and $\tan \beta =0.5$ or 2. }
\label{gghaa}
\end{figure}

\begin{figure}
\centerline{\hbox{ \epsfig{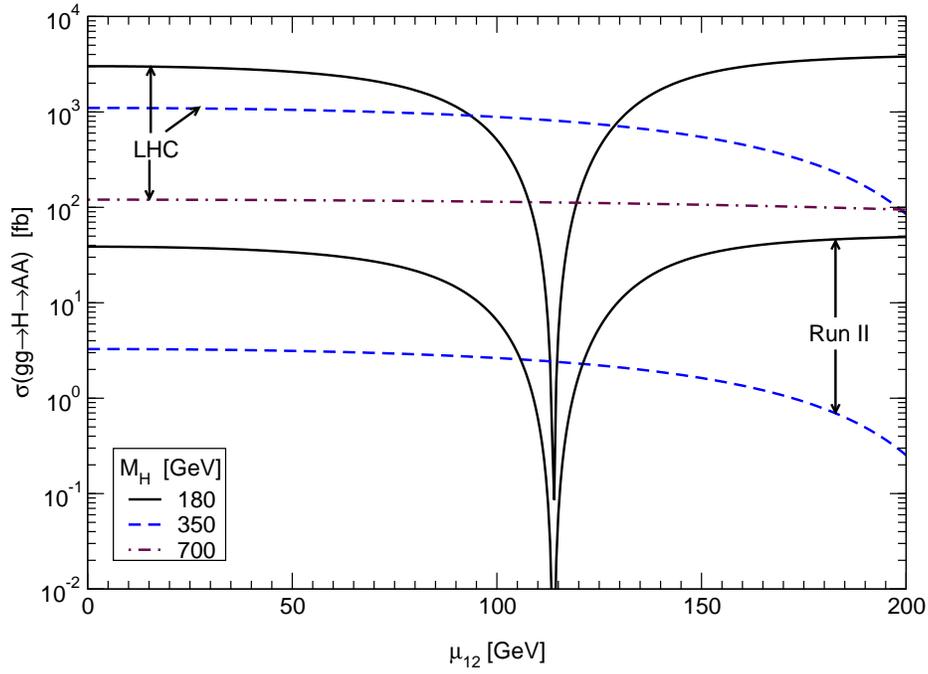}}}
\caption{ The $p{\bar p},pp(gg)  \to H \to AA$ production cross
section as a function of $\mu_{12}$ at the Tevatron Run-2 with
$\sqrt{S}=2$ TeV and the LHC with $\sqrt{S}=14$ TeV, for a few
values of $M_H$, with $M_A=0.2$ GeV and $\tan \beta =2$. }
\label{gghhaa}
\end{figure}

\begin{figure}[htb]
\centerline{\hbox{ \epsfig{figure=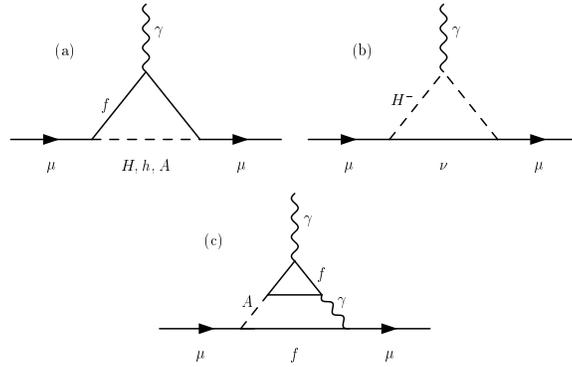,width=3in}}}
\caption{Contribution from the THDM to the anomalous magnetic
moment of muon: neutral Higgs bosons (a), charged Higgs boson (b),
and the leading two-loop contribution from the CP-odd scalar (c).}
\label{magmom}
\end{figure}

The partonic process ${\bar q}q^{\prime} \to W^+AA$ receives
contributions from just five diagrams if the quark masses are
neglected. These diagrams can be obtained from those contributing
to the process ${\bar f} f \to ZAA$, cf. Fig.~\ref{diagffzaa}.
After a few changes, the above results can also be easily
translated to obtain the respective squared amplitude. First of
all, in equation (\ref{amp}) we have $g_L=V_{qq^{\prime}}/2$ and
$g_R=0$, together with the appropriate change of notation
regarding the Dirac spinors. $V_{qq^{\prime}}$ is the CKM mixing
matrix element. Secondly, in equation (\ref{sqamp}) $\chi^2$ is
now given by

\begin{equation}
\chi^2=\frac{|V_{qq^{\prime}}|^2}{2},
\end{equation}

\noindent whereas the substitution $m_Z \to m_W$ must be done
everywhere. Finally, the $\xi_i$ form factors are defined now as

\begin{equation}
\xi_i \equiv \xi_i^W= \xi_i^h+\xi_i^H+\xi_i^{H^+},
 \label{newxi}
\end{equation}

\noindent where $\xi_0^{\phi}$ are the same as those in
(\ref{xi0}), and

\begin{equation}
\xi_{1,2}^{H^+}=\frac{1}{(k+k_{1,2})^2-M_{H^+}^2}.
\end{equation}

\noindent Again, in the resonance region, the charged Higgs boson
propagator acquires an imaginary part.

%%%%%%%%%%%%%%%%%%%%%%%%%%%%%%%%%%%%%%%%%%%%%%%%%
%%%%%%%%%%%%%%%%%%%%%%%%%%%%%%%%%%%%%%%%%%%%%%%%%

\begin{figure}
\centerline{\hbox{ \epsfig{figure=tbboundiol.eps,height=2.5in}}}
\caption{The regions (above the curves) in the $\tan\beta$ versus
$M_A$ plane of a type-I THDM allowed by the $a_\mu$ data at the
95\% CL, based on a one-loop calculation. There is no allowed
region in this range of parameters according to the DH
calculation. } \label{gm2tbma-one-1}
\end{figure}

\begin{figure}
\centerline{\hbox{ \epsfig{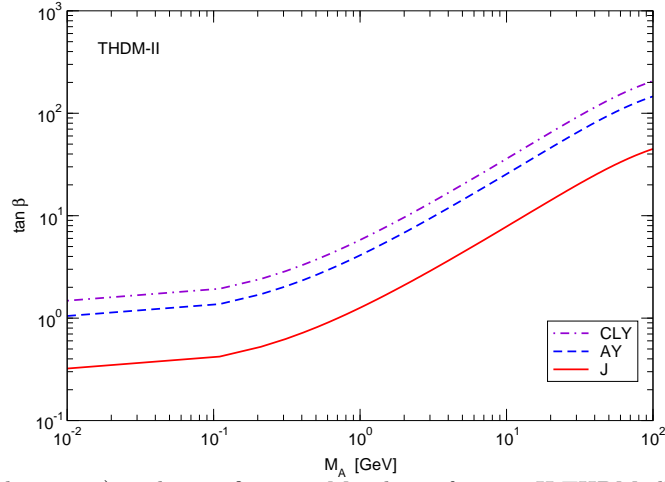}}}
\caption{The regions (below the curves) in the $\tan\beta$ versus
$M_A$ plane of a type-II THDM allowed by the $a_\mu$ data at the
95\% CL, based on a one-loop calculation. There is no allowed
region in this range of parameters according to the DH
calculation. } \label{gm2tbma-one-2}
\end{figure}

\begin{figure}
\centerline{\hbox{ \epsfig{figure=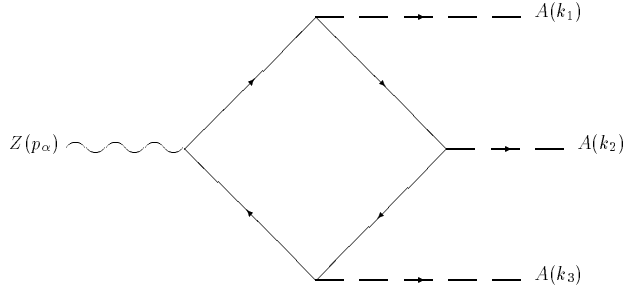,height=1.5in}}}
\caption{Representative fermion loop diagram to the $Z \to A A A$
decay.} \label{diag}
\end{figure}

\begin{table}[h]
\begin{center}
\begin{tabular}{|c||c|c||}
\hline
Constraint & Type-I THDM & Type-II THDM \\
\hline
$(g-2)_\mu$ & $\tan \beta > 0.4 $ & $\tan \beta < 2.6 $ \\
\hline {\tiny ($\tan \beta >1$)} $b\to s \gamma$ &
$M_{H^+}>100$ GeV & $M_{H^+}>200$ GeV   \\
\hline {\tiny ($0.5< \tan \beta <1$)} $b\to s \gamma$ &
--- & $M_{H^+}>200\, -350$ GeV  \\
\hline {\tiny ($0.6< \tan \beta <1$)} $R_b$ &
$M_{H^+}>200\, -600$ GeV &  $M_{H^+}>200\, -600$ GeV \\
\hline {\tiny ($\sin (\beta -\alpha)=1$)} $\Delta \rho$ &
$M_{H}\sim M_{H^+}$ &  $M_H\sim M_{H^+}$ \\
\hline {\tiny ($\sin (\beta -\alpha)=0.8$)} $\Delta \rho$ &
$M_{H}\sim 1.2 \, M_{H^+}$ &  $M_H\sim 1.2 \, M_{H^+}$ \\
\hline {\tiny ($\sin (\beta -\alpha)=0.5$)} $\Delta \rho$ &
$M_{H}\sim 1.7 \, M_{H^+}$ &  $M_H\sim 1.7 \, M_{H^+}$ \\
\hline \hline
\end{tabular}
\end{center}
\vskip 0.08in \caption{ Constraints from the low energy data for
type-I and type-II models, with $M_A=0.2$ GeV. Here, we use the
CLY calculation for the SM prediction of $a_\mu$. When $\sin
(\beta-\alpha)$ is fixed to be 1, there is no $M_h$ dependence in
$\rho$, otherwise, we assume $M_h=110$ GeV.}
\label{allconstraints}
\end{table}

\begin{table}[htb]
\begin{center}
\begin{tabular}{|c|c|c|c|c|}
\hline
THDM&$C_H $&$C_h$&$C_A$&$C_{H^\pm}$\\
\hline I&$\sin \alpha/\sin\beta$&$\cos
\alpha/\sin\beta$&$ \cot\beta$&$ \cot\beta$\\
\hline II&$\cos \alpha/\cos\beta$&$-\sin
\alpha/\cos\beta$&$- \tan\beta$&$-\tan\beta$\\
\hline
\end{tabular}
\end{center}
\caption{Higgs scalars-muon couplings in the THDM.}
\label{couplings}
\end{table}


\begin{references}

\bibitem{dobrescu} B. Dobrescu, G. Landsberg, and
K. Matchev, Phys. Rev. D {\bf 63}, 075003 (2001).

\bibitem{composite} D. Dobrescu, Phys. Rev. D {\bf 63}, 015004 (2001).

\bibitem{hunter} J. F. Gunion, H. E. Haber, G. Kane,
and S. Dawson, {\it The Higgs Hunter's Guide}, (Addison Wesley,
Reading, MA, 1996);  hep-ph/9302272 (E).

\bibitem{alephc} R. Barate {\it et al.}, Phys. Lett. B {\bf 487},
253 (2000); M. Acciari {\it et al.}, {\it ibid.} {\bf 466}, 71
(1999).

\bibitem{yukawa} R. Barate {\it et al.}, {\it Search for a Light Higgs Boson in the
Yukawa Process}, Contribution to ICHEP96 Warsaw, Pol. 25-31 July
1996, PA13-027; P. Abreu {\it et al.}, CERN-OPEN-99-385.

\bibitem{bnlgm2} H. N. Brown, {\it et. al.}, Phys. Rev. Lett. {\bf 86}, 2227  (2001).

\bibitem{marciano} A. Csarnecki and W. J. Marciano, hep-ph/0102122.

\bibitem{haberdedes} A. Dedes and H. E. Haber, hep-ph/0102297.

\bibitem{Yndurain} F. J. Yndur\'ain, hep-ph/0102312.

\bibitem{davier} M. Davier and A. H\"{o}cker,
Phys. Lett. B {\bf 435}, 427 (1998).

\bibitem{ay} K. Adel and F. J. Yndur\'ain,
hep-ph/9509378.

\bibitem{cly} A. Casas, C. L\'opez and F. J. Yndur\'ain,
Phys. Rev. D {\bf 32}, 736 (1985).

\bibitem{oneloop} W. A. Bardeen, R. Gastmans, and B. Lautrup, Nucl. Phys. {\bf B46},
319 (1972); J. P. Leveille, {\it ibid.} {\bf B137}, 63 (1978); H.
E. Haber, G. L. Kane, and T. Sterling, {\it ibid.} {\bf B161}, 493
(1979); E. D. Carlson, S. L. Glashow, and U. Sarid, {\it ibid.}
{\bf B309}, 597 (1988); J. R. Primack and H. R. Quinn, Phys. Rev.
D {\bf 6}, 3171 (1972).

\bibitem{gm2mssm} R. Barbieri and L. Maiani, Phys. Lett. B {\bf 117}, 203 (1982);
J. Ellis, J. S. Hagelin, and D. V. Nanopoulos, {\it ibid.} {\bf
116}, 283 (1982); D. A. Kosower, L. M. Krauss, and N. Sakai, {\it
ibid.} {\bf 133}, 305 (1983); M. Carena, G. F. Giudice and C. E.
M. Wagner, {\it ibid.} {\bf 390}, 234 (1997); G.-C. Cho, K.
Hagiwara, and M. Hayakawa, {\it ibid.} {\bf 478}, 231 (2000); T.
C. Yuan, R. Arnowitt, A. H. Chamseddine and P. Nath, Z. Phys. C
{\bf 26}, 407 (1984); A. Barroso, M. C. Bento, G. C. Branco, and
J. C. Romao, Nucl. Phys. {\bf B250}, 295 (1985); I. Vendramin,
Nuovo Cimento 101 {\bf A}, 731 (1989); J. A. Grifols and A.
Mendez, Phys. Rev. D {\bf 26}, 1809 (1982); J. L. Lopez, D. V.
Nanopoulos and X. Wang, {\it ibid.} {\bf 49}, 366 (1994); U.
Chattopadhyay and P. Nath, {\it ibid.} {\bf 53}, 1648 (1996); T.
Moroi, {\it ibid.} {\bf 53}, 6565 (1996); {\bf 56}, 4424(E)
(1997); W. Hollik, J. I. Illana, C. Schappacher, D. St\"ockinger,
and S. Rigolin, hep-ph/9808408; R. A. Diaz, R. Martinez and J.
-Alexis Rodriguez, hep-ph/0103050.

\bibitem{darwin} D. Chang, W. -F. Chang, C.-H. Chou, and
W.-Y. Keung, Phys. Rev. D {\bf 63}, 091301 (2001).

\bibitem{maria} M. Krawczyk and J. Zochowski,
Phys. Rev. D {\bf 55}, 6968 (1997).

\bibitem{CUSB} P. Franzini {\it et al.}, Phys. Rev. D {\bf 35}, 2883 (1987). J.
Lee-Franzini, in {\it Proceedings of the XXIV International
Conference on High Energy Physics}, Munich, Germany, 1988, edited
by R. Koffhaus and J. H. K\"uhn, Springer-Verlag, Berlin, 1989 p.
1432.

\bibitem{neubert} A. L. Kagan and M. Neubert,
Eur. Phys. J. C {\bf 7}, 5 (1999).


\bibitem{borzumati} F. M. Borzumati and C. Greub,
Phys. Rev. D {\bf 59}, 057501 (1999); {\bf 58}, 074004 (1998);
hep-ph/9810240; see also, M. Ciuchini, G. Degrassi, P. Gambino,
and G. F. Giudice, Nucl. Phys. {\bf B527}, 21 (1998); P. Gambino,
Nucl. Phys. Proc. Suppl. {\bf 86}, 499 (2000); CERN-TH-2001-029.


\bibitem{cleo} S. Ahmed  {\it et al.}, hep-ex/9908022;
T. E. Coan, hep-ph/0011098.

\bibitem{bsgaleph} R. Barate {\it et al.}, Phys. Lett. B {\bf
429}, 169 (1998).

\bibitem{belle} K. Abe {\it et al.}, hep-ex/0103042.

\bibitem{others} H. E. Haber, in {\it Perspectives on Higgs Physics},
edited by G.L. Kane, World Scientific, Singapore, 1993, p. 79; A.
Kundu and B. Mukhopadhyaya, Int. J. Mod. Phys. A {\bf 11}, 5221
(1996).  A. Denner, R. J. Guth, W. Hollik, and J. H. Kuhn, Z.
Phys. C {\bf 51}, 695 (1991). A. Djouadi, {\it et al.}, Nucl.
Phys. {\bf B349}, 48 (1991); M. Boulware and D. Finnell, Phys.
Rev. D {\bf 44}, 2054 (1991). A. K. Grant, {\it ibid.} {\bf 51},
207 (1995).

\bibitem{haber} H. E. Haber and H. E. Logan,
Phys. Rev. D {\bf 62}, 015011 (2000).

\bibitem{logan} H. E. Logan, {\it Radiative
corrections to the $Z\to b\ov b$ vertex and constraints on
extended Higgs sectors}, PhD Thesis, hep-ph/9906332.

\bibitem{s_data}D. E. Groom, {\it et al.} Eur. Phys. J. C. {\bf 15} 1 (2000).

\bibitem{s_exp}P. H.  Chankowski, M. Krawczyk, and J. Zochowski
Eur. Phys. J. C {\bf 11}, 661 (1999); G.-C. Cho and K. Hagiwara,
Nuc. Phys. {\bf B574}, 623 (2000); K. Hagiwara, D. Haidt, C. S.
Kim, and S. Matsumoto, Z. Phys. C {\bf 64}, 559 (1994); {\bf 68}
352(E) (1995). T. Inami, C. S. Lim, and A. Yamada, Mod. Phys.
Lett. A {\bf 7}, 2789 (1992).

\bibitem{field} J. Field, Mod. Phys. Lett. A {\bf 13}, 1937 (1998).

\bibitem{limits} LEP Electroweak Working Group,
hep-ex/0103048.

\bibitem{shinya} S. Kanemura, T. Kasai, and Y. Okada,
Phys. Lett. B {\bf 471}, 182 (1999).

\bibitem{akeroyd} A. G. Akeroyd, A. Arhrib, and E. Naimi,
hep-ph/0002288; A. G. Akeroyd, Nucl. Phys. {\bf B544}, 557 (1999).

\bibitem{lepfermiop} A. Rosca, hep-ex/0011082; M. Aciarri {\it et al.},
Phys. Lett. B {\bf 489}, 115 (2000); L3 Note 2526 (2000); R.
Barate {\it et al.}, Phys. Lett. B {\bf 487}, 241 (2000);  P.
Abreu {\it et al.}, {\it ibid.}  {\bf 507}, 89 (2001).

\bibitem{lepsmh} R. Barate {\it et al.}, hep-ex/0011045; P. Abreu {\it et al.},
hep-ex/0011043;  see also, J. Ellis, hep-ex/0011086.

\bibitem{ggh}
H. Georgi, S. Glashow, M. Machacek, and D. V. Nanopoulos, Phys.
Rev. Lett. {\bf 40} (1978) 692; A. Krause, T. Plehn, M. Spira, and
P. M. Zerwas, Nucl. Phys. {\bf B519}, 85 (1998); A. Djouadi, M.
Spira, and P. M. Zerwas, Phys. Lett. B {\bf 264}, 440 (1991); S.
Dawson, Nucl. Phys. {\bf B359}, 283 (1991); D. Graudenz, M. Spira,
and P. M. Zerwas, Phys. Rev. Lett. {\bf 70}, 1372 (1993); M.
Spira, A. Djouadi, D. Graudenz, and P. M. Zerwas, Phys. Lett. B
{\bf 318}, 347 (1993); M. Spira, A. Djouadi, D. Graudenz, and P.
M. Zerwas, Nucl. Phys. {\bf B453}, 17 (1995). M. Spira,
hep-ph/9711394.

\bibitem{hphm}E. Eichten, I. Hinchliffe, K. Lane, and C. Quigg
Rev. Mod. Phys. {\bf 56}, 579 (1984); {\it ibid.} {\bf 58}, 1065
(1986).

%references of the appendix

\bibitem{li} L.-F. Li, in {\it 14th International Warsaw Meeting on
Elementary Particle Physics}, Warsaw, Pol., 1991.

\bibitem{chang} D. Chang and W.-Y. Keung,
Phys. Rev. Lett. {\bf 77}, 3732 (1996).

\bibitem{van} J.J. van der Bij and E.W.N. Glover,
Nuc. Phys. {\bf B313}, 237 (1989).

\bibitem{add} K. Cheung, C.-H. Chou, O. C. W. Kong, hep-ph/0103183;
Maria Krawczyk, hep-ph/0103223.

\bibitem{gambino} P. Gambino and M. Misiak, hep-ph/0104034.

\end{references}
\end{document}